\documentclass[fleqn]{article}
\usepackage{longtable}
\usepackage{graphicx}

\begin {document}
\begin{flushleft}
{\LARGE
{\bf Energy levels and radiative rates  for transitions in Ti VI}
}\\

\vspace{1.5 cm}

{\bf {K M  ~Aggarwal$^1$,  F  P   ~Keenan$^1$ and A  Z  Msezane$^2$}}\\ 

\vspace*{1.0cm}

$^1$Astrophysics Research Centre, School of Mathematics and Physics, Queen's University Belfast, Belfast BT7 1NN, Northern Ireland, UK\\ 
$^2$Center for Theoretical Studies of Physical Systems, Clark Atlanta University, Atlanta, Ga 30314, USA
\vspace*{0.5 cm} 

e-mail: K.Aggarwal@qub.ac.uk \\

\vspace*{1.50cm}

Received  13 March 2013\\
Accepted for publication xx Month 2013 \\
Published 9  July 2013 \\
Online at stacks.iop.org/PhysScr/vol/number \\

\vspace*{1.5cm}

PACS Ref: 31.25 Jf, 32.70 Cs,  95.30 Ky

\vspace*{1.0 cm}

\hrule

\vspace{0.5 cm}
{\Large {\bf S}} This article has associated online supplementary data files \\
Tables 5 and 6 are available only in the electronic version at stacks.iop.org/PhysScr/vol/number

\end{flushleft}

\clearpage


\begin{abstract}

We report on calculations of energy levels, radiative rates, oscillator strengths, and line strengths for transitions
among the lowest 253 levels of the (1s$^2$2s$^2$2p$^6$) 3s$^2$3p$^5$, 3s3p$^6$,  3s$^2$3p$^4$3d,  3s3p$^5$3d,  3s$^2$3p$^3$3d$^2$,  3s$^2$3p$^4$4s, 3s$^2$3p$^4$4p and 3s$^2$3p$^4$4d  configurations of Ti VI. The general-purpose relativistic atomic structure package ({\sc grasp}) and flexible atomic code ({\sc fac}) are adopted for the calculations. Radiative rates, oscillator strengths and line strengths are reported for all electric dipole (E1), magnetic dipole (M1), electric quadrupole (E2) and magnetic
quadrupole (M2) transitions among the 253 levels, although calculations have been performed for a much larger number of levels.  Comparisons are made with existing available results and the accuracy of the data is assessed. Additionally, lifetimes for all 253 levels are listed, although comparisons with other theoretical results are limited to only 88 levels. Our energy levels are estimated to be accurate to better than 1\% (within 0.03 Ryd), whereas results for other parameters are probably accurate to  better than 20\%. A reassessment of the energy level data on the NIST website for Ti VI is suggested.

\end{abstract}

\clearpage

\section{Introduction}

Iron group elements (Sc - Zn) are becoming increasingly important in the study of astrophysical plasmas, as many of their  lines are frequently observed from different ionisation stages. These lines provide a wealth of data about the plasma characteristics, including  temperature, density and chemical composition. Additionally, iron group elements are often impurities in fusion reactors, and to estimate the power loss from the impurities, atomic data (including energy levels and  oscillator strengths or radiative decay rates) are required for many ions. The need for atomic data has become even greater with the developing  ITER project. Since there is a paucity of measured parameters, one must depend on theoretical results. Therefore,  recently we have reported atomic parameters for many ions of the iron group elements -- see for example  \cite{fe26}--\cite{adt1}  and references therein. Among Ti ions, results have already been provided for Ti XXII \cite{ti22}, Ti XXI \cite{ti21}, Ti XX \cite{ti20} and Ti XIX \cite{ti19}, and here we focus our attention on Cl-like Ti VI.

Several emission lines of Ti ions have been observed in astrophysical plasmas, as listed in the CHIANTI database at ${\tt {\verb+http://www.chiantidatabase.org+}}$. We are not aware of any astrophysical observations for Ti VI, but many emission lines are listed in the 125-525 ${\rm \AA}$ wavelength range in the {\em Atomic Line List} (v2.04) of Peter van Hoof at ${\tt {\verb+http://www.pa.uky.edu/~peter/atomic/+}}$, because these are useful in the generation of synthetic spectra. However, laboratory measurements for  lines of Ti VI were made as early as 1936 \cite{wk}. These and other later (but limited) measurements have been  compiled by Sugar and Corliss \cite{sc} and are also available at the NIST (National Institute of Standards and Technology)  website {\tt http://physics.nist.gov/PhysRefData/ASD/levels\_form.html}.

As with measurements, theoretical work on Ti VI is also limited. Combining solar and laboratory measurements for some lines of Cl-like  ions with calculations based on Hartree-Fock method, Gabriel and co-workers \cite{fg1}--\cite{fg2} listed some lines of Ti VI among the levels of the (3p$^5$) $^2$P and (3p$^4$3d) $^2$D, $^2$P, $^2$S multiplets. Subsequently,  Huang {\em et al} \cite{knh} and Fawcett \cite{bcf} reported energy levels and oscillator strengths for some transitions among the 31 levels of the 3s$^2$3p$^5$, 3s3p$^6$ and 3s$^2$3p$^4$3d  configurations. For their calculations, Huang {\em et al} adopted the multi-configuration Dirac-Fock (MCDF)  method of Desclaux \cite{jpd}   and also included {\em configuration interaction} (CI) with the additional 3s3p$^5$3d, 3p$^5$3d$^2$, 3s$^2$3p$^3$3d$^2$, 3p$^6$3d, and 3s3p$ ^4$3d$^2$ configurations. In addition, they reported A- values for the electric dipole (E1)  transitions from the levels of the (3s$^2$3p$^5$) $^2$P$^o_{3/2,1/2}$ ground configuration to higher excited levels, and for the magnetic dipole (M1) and electric quadrupole (E2) transitions between the two levels of the ground state.  However,  Fawcett used the Hartree-Fock relativistic (HFR) method of Cowan \cite{hfr} {\em and} also calculated energies for the levels of the 3s$^2$3p$^4$4s configuration.   These and other similar data are included in the NIST compilations.  Bi{\'{e}}mont and Tr{\"{a}}bert \cite{bt} also adopted the HFR method and included extensive CI, but were primarily focused on the lifetime of the  3s3p$^6$ $^2$S$_{1/2}$ level. Nevertheless, their calculated energies for the 3s$^2$3p$^5$ $^2$P$^o_{1/2}$ and 3s3p$^6$ $^2$S$_{1/2}$ levels differed with the NIST compilations by  7\% and 1\%, respectively.   

More recently, Froese Fischer {\em et al}  \cite{cff} calculated energies for the levels of the 3s$^2$3p$^5$, 3s3p$^6$,  3s$^2$3p$^4$3d and 3s$^2$3p$^4$4s configurations using their multi-configuration Hartree-Fock (MCHF) method. Similar to other researchers, they also included relativistic effects along with CI with some configurations, but  up to $n$ = 7 and $\ell$ = 4. However, their reported results are limited to only a few levels/transitions and hence are insufficient for plasma modelling. Finally,  Mohan {\em et al} \cite{mm1} have performed a comparatively larger calculation, as they have also included 57 levels of the 3s$^2$3p$^4$4s,  3s$^2$3p$^4$4p and 3s$^2$3p$^4$4d configurations, apart from the lowest 31 levels of the 3s$^2$3p$^5$, 3s3p$^6$ and  3s$^2$3p$^4$3d configurations. Furthermore, they have included an extensive CI (see Table 3 of \cite{mm1}) for the construction of wavefunctions, apart from one-body relativistic operators in the Breit-Pauli approximation. For their calculation, they adopted the CIV3 program of Hibbert  \cite{civ3}. However, as in earlier work, they also only reported A- values for E1 transitions, whereas in plasma modelling the A- values for all types of transitions, namely electric dipole (E1), electric quadrupole (E2), magnetic dipole (M1), and magnetic quadrupole (M2), are required as demonstrated by Del Zanna {\em et al}\,  \cite{del04}. Additionally, Mohan {\em et al} did {\em not} calculate energies for the levels of the 3s3p$^5$3d and 3s$^2$3p$^3$3d$^2$ configurations, although these have been included in the generation of wavefunctions. However, these two configurations give rise to 164 levels which closely interact and intermix with those of the 3s$^2$3p$^5$, 3s3p$^6$,  3s$^2$3p$^4$3d, 3s$^2$3p$^4$4s, 3s$^2$3p$^4$4p  and 3s$^2$3p$^4$4d configurations, included by  Mohan {\em et al}. These missing levels from the calculations of Mohan {\em et al} and other workers are required in plasma modelling, as they affect the construction of the synthetic spectrum as well as the calculation of lifetimes.  

Apart from the exclusion of the levels of the 3s3p$^5$3d and 3s$^2$3p$^3$3d$^2$ configurations, Mohan {\em et al} \, \cite{mm1} stressed the need for further work on Ti VI, because there is a {\em discrepancy} in energy levels and their orderings with the NIST compilations. Therefore,  the {\em aim} of the present paper is not only to improve upon the calculations of Mohan {\em et al} but also to report a {\em complete} set of results, among the lowest 253 levels of  Ti VI, which can be confidently applied in plasma modelling. These levels include all those of the 3s$^2$3p$^5$, 3s3p$^6$,  3s$^2$3p$^4$3d,  3s3p$^5$3d, 3s$^2$3p$^3$3d$^2$,  3s$^2$3p$^4$4s, 3s$^2$3p$^4$4p and 3s$^2$3p$^4$4d  configurations, but only a few of 3s3p$^4$3d$^2$. As well as  energy levels, we report  radiative rates for all E1, E2, M1 and M2 transitions among these 253 levels  of Ti VI, and provide theoretical lifetimes for all levels. Comparisons are made with earlier available theoretical results and the accuracy of the data is assessed. 


\section{Energy levels}

For our calculations we have adopted the {\sc grasp} (general-purpose relativistic atomic structure package) code to generate the wavefunctions. This code was originally developed  by Grant {\em et al} \cite{grasp0}, but the name {\sc grasp} was first assigned by Dyall {\em et al} \cite{grasp1} in the modified version.  It has been further  revised by several workers under the names  {\sc grasp92} \cite{grasp2} and {\sc grasp2k}  \cite{grasp2k}. However, the version adopted here is based on \cite{grasp0} and has been revised by one of the authors (Dr P H Norrington), who prefers to refer to it as  {\sc grasp0}.  This version contains most of the modifications undertaken in the other revised codes and is available on the website  {\tt http://web.am.qub.ac.uk/DARC/}. A particularly useful feature of this version is that it also has provisions for listing the $LSJ$ designations of the levels/configurations, apart from the usual $jj$ nomenclature of the relativistic codes. Another useful advantage of this version is that its output can be directly linked to the collisional code (i.e. the {\em Dirac atomic R-matrix code}, DARC), although calculations of the collisional data is not the subject of this paper.

{\sc grasp} is a fully relativistic code, and is based on the $jj$ coupling scheme. Further higher order relativistic terms arising from the Breit interaction and QED (quantum electrodynamics) effects (vacuum polarisation and Lamb shift)  have also been included in the same way as in the original version described in \cite{grasp0} {\em and} \cite{graspt2}. Additionally, we have used the option of {\em extended average level} (EAL),  in which a weighted (proportional to 2$j$+1) trace of the Hamiltonian matrix is minimised. This produces a compromise set of orbitals describing closely lying states with moderate accuracy, and generally yields results comparable to other options, such as {\em average level} (AL), as noted by Aggarwal {\em et al}  for several ions of Kr \cite{kr} and Xe \cite{xe}.  

Although  Ti VI is moderately heavy ($Z$ = 22) and 5 times ionised, {\em configuration interaction} (CI) is still very {\em important} for an accurate determination of energy levels. For this reason most earlier workers have included CI with additional configurations, although their calculations have generally been confined to the lowest 31 levels of the 3s$^2$3p$^5$, 3s3p$^6$ and  3s$^2$3p$^4$3d configurations. Following some tests with a number of $n$ = 3, 4 and 5 configurations, we  have also arrived at the same conclusion that an elaborate CI needs to be included to achieve a better accuracy in the determination of energy levels. Therefore, we have performed a series of calculations  with increasing amount of CI, but focus only on three, namely (i) GRASP1, which includes 60 levels among the 3s$^2$3p$^5$, 3s3p$^6$,  3s$^2$3p$^4$3d,  3s$^2$3p$^4$4s and 3s$^2$3p$^4$4p configurations, (ii) GRASP2, with an additional 508 levels of the 3s$^2$3p$^4$4d, 3s$^2$3p$^4$4f, 3s3p$^5$3d, 3s$^2$3p$^3$3d$^2$, 3s3p$^4$3d$^2$, 3p$^6$3d, 3p$^6$4s and 3s3p$^5$4$\ell$  configurations, and finally (iii) GRASP3, with another 3181 levels of the 3s$^2$3p$^4$5$\ell$, 3s3p$^5$5$\ell$, 3s$^2$3p$^2$3d$^3$, 3s3p$^3$3d$^3$, 3p$^5$3d$^2$,  3s$^2$3p3d$^4$, 3s3p$^2$3d$^4$,  3p$^6$4p/d/f and 3p$^3$3d4$\ell$  configurations, i.e. 3749 levels in total among the above listed 38 configurations. Before we discuss our results, we note that calculations have also been performed which include   configurations such as 3s3p3d$^5$, 3s$^2$3d$^5$, 3p$^2$3d$^5$, 3s3d$^6$, 3p3d$^6$ and 3d$^7$. However, these are not discussed here because their impact on the levels considered  is insignificant.

\subsection{$n$ = 3 levels}

In Table 1 we list our calculated energies with the  {\sc grasp} code for the lowest 31 levels of the 3s$^2$3p$^5$, 3s3p$^6$ and  3s$^2$3p$^4$3d configurations. Results from all three calculations with the {\sc grasp} code described above  are listed here, and  include the Breit and QED corrections. However, we  stress  that the Breit contribution is up to 0.005 Ryd, depending on the level,  whereas that of QED is negligible. Also included in this table are the energies obtained by Fawcett \cite{bcf}, Froese Fischer {\em et al} \cite{cff} and Mohan {\em et al} \cite{mm1} from the HFR, MCHF and  CIV3  codes, respectively. The corresponding energies of Huang {\em et al}\, \cite{knh}  with the MCDF code are excluded, because they included only limited CI among the  3s$^2$3p$^5$, 3s3p$^6$, 3s$^2$3p$^4$3d,  3s3p$^5$3d, 3p$^5$3d$^2$, 3s$^2$3p$^3$3d$^2$, 3p$^6$3d and 3s3p$ ^4$3d$^2$ configurations. As a result, their energies are the highest among all other calculations, and differ by up to 0.2 Ryd. For the 3s$^2$3p$^5$  $^2$P$^o_{1/2}$ level, their energy (0.04873 Ryd) is lower than from most other calculations and is only marginally higher than the  MCHF result of 0.04614 Ryd.
However, the energies compiled by NIST  are  included in the table, although for most of the levels (excepting lowest three) are not based on measurements, as noted in section 1. A major problem with the NIST listings is the {\em orderings} of the 3p$^4$($^3$P)3d $^2$P$_{1/2,3/2}$, $^2$D$_{3/2,5/2}$ and 3p$^4$($^1$D)3d $^2$P$_{1/2,3/2}$, $^2$D$_{3/2,5/2}$ levels,  which are reversed and do not agree with most of the calculations, as elaborated below.  

Mohan {\em et al}\,  \cite{mm1} have discussed the {\em discrepancy} of level orderings, which is due to level mixing (also see Fawcett \cite{bcf}), but have still preferred to follow the NIST orderings. However, we obtain the same orderings of these levels in all calculations, irrespective of the size of CI included, as seen in Table 1. In most CI calculations, a level designation is assigned based on the strength of the corresponding  eigenvector/mixing coefficient. Sometimes the levels are so highly mixed that the same eigenvector may dominate over several levels and thus an unambiguous  identification is not possible. However, this is not the case for the levels listed in Table 1, although some of these are well mixed. For example, in our GRASP3 calculations the mixing coefficients for level 9 are  0.716 and 0.664 for 3p$^4$($^1$D)3d $^2$P$_{1/2}$ and 3p$^4$($^3$P)3d $^2$P$_{1/2}$, respectively, and for level 29 are 0.711 and 0.663 for 3p$^4$($^3$P)3d $^2$P$_{1/2}$ and 3p$^4$($^1$D)3d $^2$P$_{1/2}$, respectively. Similarly,  the mixing coefficients for level 13 are  0.714 and 0.648 for 3p$^4$($^1$D)3d $^2$P$_{3/2}$ and 3p$^4$($^3$P)3d $^2$P$_{3/2}$, respectively, and for level 28 are 0.708 and 0.647 for 3p$^4$($^3$P)3d $^2$P$_{3/2}$ and 3p$^4$($^1$D)3d $^2$P$_{3/2}$, respectively. Thus the identifications for the 3p$^4$($^3$P)3d $^2$P$_{1/2,3/2}$ and 3p$^4$($^1$D)3d $^2$P$_{1/2,3/2}$ levels are clear, and so are for the 3p$^4$($^3$P)3d  $^2$D$_{3/2,5/2}$ and 3p$^4$($^1$D)3d  $^2$D$_{3/2,5/2}$ levels.

Our orderings of energy levels are also confirmed by the calculations of Fawcett \cite{bcf} and Froese Fischer {\em et al} \cite{cff}. Therefore, if the orderings of the 3p$^4$($^3$P)3d $^2$P$_{1/2,3/2}$, $^2$D$_{3/2,5/2}$ and 3p$^4$($^1$D)3d $^2$P$_{1/2,3/2}$, $^2$D$_{3/2,5/2}$ levels is reversed in the NIST listings then the discrepancy in energy levels of $\sim$ 1 Ryd reduces to only $\sim$ 0.1 Ryd.

The calculations of Fawcett \cite{bcf} and Froese Fischer {\em et al} \cite{cff} provide similar energy levels, but the MCHF energy for the 3p$^5$ $^2$P$^o_{1/2}$ level is lowest among all calculations listed in Table 1, and differs by up to 11\%. For other levels, differences between our GRASP1 and MCHF \cite{cff} energies are up to 3\%, with the MCHF energies being higher for some (e.g. 1--23), whereas for others (25--31) they are lower. As the HFR, MCHF and CIV3 energies are closer to each other for most of the levels (with the adjustment of orderings for CIV3) and differ with our GRASP1 calculations, by up to 0.2 Ryd, we discuss these further to understand the discrepancy.

CI is very important for Ti VI as already noted and hence we have performed larger calculations with 568 and 3749 levels in GRASP2 and GRASP3, and these results are also included in Table 1. However, our GRASP2 energies have increased for most of the levels, by up to 0.17 Ryd, and the discrepancies therefore have become larger. This clearly indicates that GRASP2 does not have sufficient CI and therefore a much larger calculation is required, such as GRASP3. Our GRASP3 energies are lower than those from GRASP2 for most of the levels, by up to 0.2 Ryd, and are much closer to the HFR \cite{bcf}, MCHF \cite{cff} and CIV3 \cite{mm1} results. Therefore, we investigate the effect of further CI on the energies of these lowest 31 levels.

With our available computational resources it is not feasible to experiment with much larger CI using the {\sc grasp} code than what is already included in GRASP3 calculations discussed above. However, we can perform  much larger calculations with the  {\em Flexible Atomic Code} ({\sc fac}) of Gu \cite{fac},  available from the website {\tt http://kipac-tree.stanford.edu/fac}. This is also a fully relativistic code which provides a variety of atomic parameters, and yields results for energy levels and A- values comparable to {\sc grasp}, as already shown for several other ions, see for example:  Aggarwal {\em et al} \cite{fe15} for  Mg-like ions and \cite{ti22}--\cite{ti19} for Ti ions. In addition, a clear advantage of this code is its very high efficiency which means that large calculations can be performed within a reasonable time frame of a few weeks. Thus results from {\sc fac} will be helpful in assessing the accuracy of our energy levels and radiative rates.

As with the {\sc grasp} code, we have performed a series of calculations using {\sc fac} with increasing level of CI. However, here we focus on only two calculations, namely (i) FAC1, which includes 5821 levels among the 3*7, 3s$^2$3p$^4$ 4*1, 3s$^2$3p$^4$ 5*1,  3s3p$^5$ 4*1, 3s3p$^5$ 5*1,  3p$^6$ 4*1, 3p$^6$ 5*1 and 3s$^2$3p$^3$3d 4*1 configurations, and (ii) FAC2, which includes a total of 9160 levels, the additional ones arising from the 3s$^2$3p$^4$ 6*1,  3s3p$^5$ 6*1,  3p$^6$ 6*1, 3s$^2$3p$^3$3d 5*1 and 3s$^2$3p$^3$3d 6*1 configurations. The results obtained from these two calculations are listed in Table 1. Both calculations yield the same {\em orderings} as in our present results with {\sc grasp} or the earlier ones with MCHF \cite{cff}. Furthermore, both calculations agree within 0.01 Ryd which indicates that inclusion of larger CI in FAC2 is of no significance, or energies for the lowest 31 levels of Ti VI have fully {\em converged}. The agreement between the {\sc grasp} and {\sc fac} calculations is within 0.03 Ryd ($\sim$ 1\%), which is highly satisfactory, and there is also no discrepancy with the MCHF and CIV3 energies. Thus all calculations give comparable energies in magnitude, but the orderings of the CIV3 and NIST compilations are in disagreement with our present work using {\sc grasp} and {\sc fac}, and the earlier one with HFR and MCHF. Based on these comparisons we may conclude that a reassessment of the level orderings on the NIST website is desirable. Finally, we note that larger calculations with up to 18,459 levels (FAC3) have also been performed, but the energies obtained for the 31 levels of Table 1 are negligibly different with those from FAC2. This is because the additional levels arise from the 3s$^2$3p$^3$ 4*2, 3s$^2$3p$^3$ 5*2, 3s$^2$3p$^3$ 4*1 5*1, 3p$^5$ 4*2 and 3p$^5$3d 4*1 configurations, and their energies are well above the lowest 31 or the maximum 253 levels considered in this paper.

\subsection{3p$^4$4s and 3p$^4$4p levels}

In Table 2 we compare energies for the levels of the 3s$^2$3p$^4$4s and 3s$^2$3p$^4$4p configurations of Ti VI. Included in this table are energies from our calculations with the {\sc grasp} (GRASP1 and GRASP3) and {\sc fac} (FAC1 and FAC2) codes,  plus the earlier CIV3 results of Mohan {\em et al} \cite{mm1}. Energies for the levels of the 3s$^2$3p$^4$4s configuration are also available on the NIST website as well as from MCHF \cite{cff}, and are included in Table 2. As for the lowest 31 levels of Table 1, agreement between our GRASP1 calculations and the NIST compilations and MCHF energies is highly satisfactory, but the corresponding results with extensive CI (i.e. GRASP3 and FAC1/FAC2), are higher by up to 0.25 Ryd for almost all levels. This clearly indicates the importance of CI in the determination of energy levels for Ti VI. The CIV3 energies of Mohan {\em et al} are in between those of NIST and GRASP3 (and FAC1/FAC2). This may be indicative of  neglecting a few important configurations in their calculations, particularly 3s3p$^5$4p and 3s$^2$3p$^4$5g, whose levels lie in the 6.9--8.8 Ryd range and thus have the potential for affecting the energies of the 3s$^2$3p$^4$4s levels.

For levels of the 3s$^2$3p$^4$4p configuration, the only  results available for comparison are those of Mohan {\em et al}\,  \cite{mm1} from the CIV3 code. Their energies agree closely with our GRASP1 calculations, but are underestimated by up to 0.25 Ryd in comparison with our more elaborate GRASP3 and FAC1/FAC2  results, perhaps for the same reason as for the levels of the 3s$^2$3p$^4$4s configuration.  Since there is convergence in energies between the FAC1 and FAC2 calculations and there are also no significant differences with GRASP3, we have confidence in our results. Furthermore, based on the comparisons shown in Table 2 and discussed here we may conclude that energies for the levels of the 3s$^2$3p$^4$4s   configuration listed on the NIST website are underestimated by up to 0.25 Ryd for most of the levels, and therefore a reassessment of their data is desirable. 

\subsection{3p$^4$4d levels}

In Table 3 we compare our energies with the {\sc grasp} and {\sc fac} codes for the levels of the  3s$^2$3p$^4$4d configuration with those of NIST and CIV3 \cite{mm1}. As for other levels, agreement between our GRASP1 and NIST energies is satisfactory, but these are underestimated by up to 0.25 Ryd, because of the inclusion of limited CI. The energies and orderings in the FAC1 and FAC2 calculations are the same and confirm, once again, that additional CI included in FAC2 calculations is of no significance. Similarly, for levels of the 3s$^2$3p$^4$4d configuration there are no discrepancies with our results from GRASP3 and those of Mohan {\em et al}\,  \cite{mm1}. Thus all  calculations with extensive CI produce comparable energies.  However, Mohan {\em et al}  have not reported energies for the 3s$^2$3p$^4$($^3$P)4d $^4$D$_{7/2,5/2,3/2,1/2}$ levels, and many levels are missing from the NIST compilations. Based on the comparisons shown in Table 3, we may conclude that the energies for levels of the  3s$^2$3p$^4$4d configuration of Ti VI compiled by NIST are underestimated, whereas those of the present calculations with the {\sc grasp} and {\sc fac} codes, or the earlier one with CIV3,  are assessed to be accurate to better than 1\%.

\subsection{3p$^4$4f levels}

For the levels of the 3s$^2$3p$^4$4f configuration there are no results available in the literature with which to compare. Therefore, as for other levels, we have performed our calculations with the {\sc grasp} and {\sc fac} codes, with increasing amount of CI as stated in section 2. This configuration generates 30 levels which are listed in Table 4. Energies obtained in our GRASP1 calculations are lower than those from GRASP3, FAC1 and FAC2, by up to 0.5 Ryd -- see for example, the 3s$^2$3p$^4$($^1$D)4f $^2$P$^o_{1/2,3/2}$ levels. This is clearly due to the limited CI included in the GRASP1 calculations. Discrepancies between the GRASP3 and FAC1 (or FAC2) calculations are lower than 0.08 Ryd, which is highly satisfactory. However, for some levels the orderings are slightly different between GRASP3 and FAC1, but both calculations provide comparable results to an accuracy of better than 1\%.

\subsection{Lowest 253 levels}

In Table 5 we list our final energies, in increasing order,  obtained using the {\sc grasp} code with CI among 38 configurations listed in  section 2, which correspond to the GRASP3 calculations. These configurations generate 3749 levels, but for conciseness energies are listed only for the lowest 253 levels, which include all levels of the 3s$^2$3p$^5$, 3s3p$^6$,  3s$^2$3p$^4$3d,  3s3p$^5$3d, 3s$^2$3p$^3$3d$^2$,  3s$^2$3p$^4$4s, 3s$^2$3p$^4$4p and 3s$^2$3p$^4$4d  configurations, but only a few of 3s3p$^4$3d$^2$.  Levels of the 3s$^2$3p$^4$4f  configuration are deliberately excluded. Although this configuration generates only 30 levels, their inclusion  will span over 371, i.e. an additional 118 levels. However, energies for these levels have already been provided in Table 4. Furthermore, data corresponding to all calculations for any desired number of levels up to 18,459 can be obtained on request from the first author (K.Aggarwal@qub.ac.uk).

Although calculations with the {\sc fac} code have been performed with the inclusion of larger CI, energies obtained with the {\sc grasp} code alone are listed in Table 5. This is partly because both codes provide energies with comparable accuracy as demonstrated and discussed in sections 2.1 to 2.4, but mainly because $LSJ$ designations of the levels are also produced  in the {\sc grasp} code. For a majority of users these designations are more familiar and hence preferable. However, we note that the $LSJ$ designations provided in this table are not always unique, because some of the levels are highly mixed with others, mostly from the same but sometimes with other configurations. This has also been discussed by Mohan {\em et al} \cite{mm1}. Therefore, care has been taken to provide the most appropriate designation of a level/configuration, but a redesignation of  these cannot be ruled out in a few cases. 

Among the 253 levels listed in Table 5, comparisons with other available work is possible only for the 88 of the 3s$^2$3p$^5$, 3s3p$^6$,  3s$^2$3p$^4$3d, 3s$^2$3p$^4$4s, 3s$^2$3p$^4$4p and 3s$^2$3p$^4$4d configurations, for which extensive comparisons have been shown in Tables 1--4. Based on these comparisons it is concluded that CI is very important for the energy levels of Ti VI. Since the earlier calculations of Huang {\em et al} \cite{knh}, Fawcett \cite{bcf} and Froese-Fischer {\em et al} \cite{cff} include rather limited CI, their energy levels are not as accurate as those listed in Table 5, and are underestimated by up to 0.5 Ryd. Similarly, the energy levels compiled by NIST are not only underestimated but also have reverse orderings for the 3p$^4$($^3$P)3d $^2$P$_{1/2,3/2}$, $^2$D$_{3/2,5/2}$ and 3p$^4$($^1$D)3d $^2$P$_{1/2,3/2}$, $^2$D$_{3/2,5/2}$ levels. However, the earlier results of Mohan {\em et al} \cite{mm1} with the CIV3 code \cite{civ3}, although limited in scope, are comparatively more accurate, except for the levels of the 3s$^2$3p$^4$4s and 3s$^2$3p$^4$4p configurations, because they too have included larger CI in their calculations. Finally, based on our calculations with two independent codes, namely {\sc grasp} and {\sc fac}, with increasing amount of CI, and comparisons made with earlier (mostly theoretical) results, our energy levels listed in Table 5 are assessed to be accurate to better than 1\%.

\section{Radiative rates}

The absorption oscillator strength ($f_{ij}$) and radiative rate A$_{ji}$ (in s$^{-1}$) for a transition $i \to j$ are related by the following expression:

\begin{equation}
f_{ij} = \frac{mc}{8{\pi}^2{e^2}}{\lambda_{ji}}^2 \frac{{\omega}_j}{{\omega}_i}A_{ji}
 = 1.49 \times 10^{-16} \lambda^2_{ji} (\omega_j/\omega_i) A_{ji}
\end{equation}
where $m$ and $e$ are the electron mass and charge, respectively, $c$ is the velocity of light, 
$\lambda_{ji}$ is the transition energy/wavelength in $\rm \AA$, and $\omega_i$ and $\omega_j$ are the statistical weights of the lower $i$ and upper $j$ levels, respectively.
Similarly, the oscillator strength $f_{ij}$ (dimensionless) and the line strength $S$ (in atomic unit, 1 a.u. = 6.460$\times$10$^{-36}$ cm$^2$ esu$^2$) are related by the 
following standard equations:

\begin{flushleft}
for the electric dipole (E1) transitions: 
\end{flushleft} 
\begin{equation}
A_{ji} = \frac{2.0261\times{10^{18}}}{{{\omega}_j}\lambda^3_{ji}} S \hspace*{1.0 cm} {\rm and} \hspace*{1.0 cm} 
f_{ij} = \frac{303.75}{\lambda_{ji}\omega_i} S, \\
\end{equation}
\begin{flushleft}
for the magnetic dipole (M1) transitions:  
\end{flushleft}
\begin{equation}
A_{ji} = \frac{2.6974\times{10^{13}}}{{{\omega}_j}\lambda^3_{ji}} S \hspace*{1.0 cm} {\rm and} \hspace*{1.0 cm}
f_{ij} = \frac{4.044\times{10^{-3}}}{\lambda_{ji}\omega_i} S, \\
\end{equation}
\begin{flushleft}
for the electric quadrupole (E2) transitions: 
\end{flushleft}
\begin{equation}
A_{ji} = \frac{1.1199\times{10^{18}}}{{{\omega}_j}\lambda^5_{ji}} S \hspace*{1.0 cm} {\rm and} \hspace*{1.0 cm}
f_{ij} = \frac{167.89}{\lambda^3_{ji}\omega_i} S, 
\end{equation}

\begin{flushleft}
and for the magnetic quadrupole (M2) transitions: 
\end{flushleft}
\begin{equation}
A_{ji} = \frac{1.4910\times{10^{13}}}{{{\omega}_j}\lambda^5_{ji}} S \hspace*{1.0 cm} {\rm and} \hspace*{1.0 cm}
f_{ij} = \frac{2.236\times{10^{-3}}}{\lambda^3_{ji}\omega_i} S. \\
\end{equation}

The A- and f- values have been calculated in both Babushkin and Coulomb gauges which are  equivalent to the length and velocity forms in the non-relativistic nomenclature. However, the results are presented here in the length form alone because the velocity form requires the inclusion of negative energy states, which are not included,  and hence those results are considered to be comparatively less accurate.   In Table 6 we present transition energies ($\Delta$E$_{ij}$ in $\rm \AA$), radiative rates (A$_{ji}$ in s$^{-1}$), oscillator strengths ($f_{ij}$, dimensionless), and line strengths ($S$ in a.u.) for all 7604 electric dipole (E1) transitions among the lowest 253 levels of Ti VI. The {\em indices} used to represent the lower and upper levels of a transition have already been defined in Table 5. Also, in calculating the  above parameters we have used the Breit and QED corrected theoretical energies/wavelengths as listed in Table 5. However, only A-values are included in Table 6 for the 13,879 electric quadrupole (E2), 9954  magnetic dipole (M1), and 10,411 magnetic quadrupole (M2) transitions. Corresponding results for f- or S- values can be easily obtained by using Eqs. (1-5). 

Measurements of wavelengths are available for two transitions, namely 3s$^2$3p$^5$ $^2$P$^o_{3/2}$ -- 3s$^2$3p$^5$ $^2$P$^o_{1/2}$ and 3s$^2$3p$^5$ $^2$P$^o_{3/2}$ -- 3s3p$^6$ $^2$S$_{1/2}$, i.e. transitions 1--3 and 2--3. The corresponding values of $\lambda$ for these transitions, measured by Weissberg  and Kruger \cite{wk} and Svensson \cite{las}, are 508.6 and 524.1 $\rm \AA$, respectively. These measurement agree within 3 $\rm \AA$ with our corresponding results of 511.9 and 527.5 $\rm \AA$ for the 1--3 and 2--3 transitions, respectively.

In Table 7 we compare our oscillator strengths for transitions among the lowest 31 levels  from all three calculations with {\sc grasp} (GRASP1, GRASP2 and GRASP3) and two with {\sc fac} (FAC1 and FAC2), with those of Froese-Fischer {\em et al} \cite{cff} and Mohan {\em et al} \cite{mm1} from the MCHF and CIV3 codes, respectively. For strong transitions (f $\ge$ 0.01) there is comparatively good agreement among all calculations, except those from GRASP1, although there are differences of up to 50\% for a few, such as 1--31 and 2--28. Since the GRASP1 calculations include only limited CI, differences with other results are up to a factor of two -- see for example, transitions 1--27/29/30/31 and 2--28/29. However, the effect of CI is more apparent for the weaker transitions, because differences among the various calculations (excluding GRASP1) are up to an order of magnitude -- see, for example transitions 1--12 and 1--16. Our GRASP2 calculations particularly illustrate the inadequacy of CI as may be noted for the 1--12 and 1--26 transitions, for which the f- values differ by up to three orders of magnitude! However, there is very good agreement among our GRASP3, FAC1 and FAC2 calculations for almost all transitions listed in Table 6, although there are also some exceptions, such as the 1--12 and 1--26 transitions. Similarly, f- values from the  MCHF   calculations differ by up to a factor of three for several transitions, such as 1--5/7/16/26 and 2--6/16. Therefore, as for energy levels, the f- values from the MCHF code are not as accurate as those calculated here, because of the limited CI. However, quite unexpectedly the  f- values from the CIV3 code are also not as accurate, because discrepancies for several transitions are up to an order of magnitude -- see, for example, 1--16/17 and 2--16. This is in spite of the fact that Mohan {\em et al}  have also included a large CI in their calculations. Before we discuss the differences further, we note that f- values for weaker transitions are generally less accurate, because mixing coefficients from several components may have an additive or cancellation effect, which affects the weaker transitions more than the strong ones. Nevertheless, a normal practice in a CIV3 calculation is to first survey all levels of a configuration and then eliminate those whose eigenvectors are below a certain limit (say $\sim$ 0.2) before performing a final run for transition rates. This exercise is undertaken to keep the calculations manageable within the limited computational resources available, and is the most likely reason for the differences in f- values between our elaborate calculations with the {\sc grasp} and {\sc fac} codes and those of Mohan {\em et al} with CIV3. Finally, for the two most important transitions of Ti VI, namely 1-3 (3p$^5$ $^2$P$^o_{3/2}$ -- 3s3p$^6$ $^2$S$_{1/2}$) and 2-3 (3p$^5$ $^2$P$^o_{1/2}$ -- 3s3p$^6$ $^2$S$_{1/2}$), two other sophisticated calculations by Bi{\'{e}}mont  and Tr{\" {a}}bert \cite{bt} and Berrington {\em et al}\,  \cite{bpw} are available. For these two transitions their f- values are 0.0228 and 0.02285, and 0.02315 and 0.02355, respectively, and  the discrepancies with our results from {\sc grasp} and {\sc fac} are less than 10\%. However, the corresponding results of  Froese Fischer {\em et al} and particularly of Mohan {\em et al}   from the MCHF and CIV3 calculations differ by up to 25\%.

A general criteria to assess the accuracy of radiative rates is to compare the length and velocity forms of the f- or A- values. However, such comparisons are only desirable, and are {\em not} a fully sufficient test to assess accuracy, as calculations based on different methods (or combinations of configurations) may give comparable f- values in the two forms, but entirely different results in magnitude. Generally, there is a good agreement between the length and velocity forms of the f- values for {\em strong} transitions (f $\ge$ 0.01), but differences  between the two can sometimes be substantial even for some very strong transitions, as demonstrated through various examples by Aggarwal {\em et al} \cite{fe15}. Nevertheless, in Table 7 we have also listed the ratio of velocity and length forms of the A- values corresponding to our GRASP3 calculations. For all strong transitions (f $\ge$ 0.01) the two forms agree within 20\%, but differences are larger for weaker transitions, such as 1--7/8 and 2--6/7/9. 

In Table 8 we compare f- values from our calculations with {\sc grasp} (GRASP1 and GRASP3) and {\sc fac} (FAC1 and FAC2) with those of Mohan {\em et al} \cite{mm1} from the CIV3 code for transitions from  3p$^5$ $^2$P$^o_{3/2,1/2}$ to levels of the (3p$^4$) 4s and 4d configurations. Our FAC1 and FAC2 calculations provide almost the same results (with only the exception of 3p$^5$ $^2$P$^o_{3/2}$ -- 3p$^4$($^3$P)4d $^4$P$_{5/2}$, for which f $\sim$ 10$^{-4}$), indicating again the redundancy of the additional CI included in FAC2. The  corresponding results from GRASP1 differ by up to two orders of magnitude for a few transitions, particularly the weaker ones (f $\le$ 10$^{-5}$) due to the insufficient inclusion of CI. However, there is no (major) discrepancy between our GRASP3 and FAC1 f- values, although the differences for a few weak transitions are up to a factor of two, and the result for the 3p$^5$ $^2$P$^o_{1/2}$ -- 3p$^4$($^3$P)4d $^4$F$_{3/2}$ transition is anomalous in GRASP3. Such anomalies for a few transitions in a large calculation are quite common irrespective of the method/code adopted. On the other hand the f- values of Mohan {\em et al} differ by a factor of two for a majority of transitions, and by an order of magnitude for a few weaker ones with f $\sim$ 10$^{-5}$. The close similarity of our results from the  {\sc grasp} and {\sc fac} codes confirm that the corresponding f- values of Mohan {\em et al} are not accurate, for the same reasons as explained above.

As in Table 7, in Table 8 also we have listed the ratio of velocity and length forms of the A- values corresponding to our GRASP3 calculations. For all strong transitions (f $\ge$ 0.01) the two forms agree within 20\%, which is highly satisfactory. However, differences are larger for some weaker transitions. Comparisons of the length and velocity forms of the A- values shown in Tables 7 and 8 are only for a few selected transitions, although they do give an indication of accuracy of our data. Similar comparisons made for all E1 transitions show that 
for almost all  strong transitions  the two forms agree to within 20\%, but differences for 224 ($\sim$3\%) of the transitions are larger (up to 50\%), and for five transitions (50--250: f = 0.026, 52--250: f = 0.012, 54--232: f=0.014, 55--233: f=0.010 and 59--236: f = 0.013), the two forms differ by  up to an order of magnitude. Therefore, on the basis of these and earlier comparisons shown in Tables 7 and 8 we may state that for a majority of the strong E1 transitions, our radiative rates are accurate to better than 20\%. However, for the weaker transitions this assessment of accuracy does not apply, because such transitions are very sensitive to mixing coefficients, and hence differing amount of CI (and methods) produce different f- values, as discussed in detail by Hibbert \cite{ah3}. This is the main reason that the two forms of f- values for some weak transitions differ significantly (by orders of magnitude), and examples include 52--151 (f  $\sim$10$^{-9}$), 26--33 (f $\sim$10$^{-8}$) and 31--119 (f $\sim$10$^{-7}$). The f-values for weak transitions may be required in plasma modelling for completeness, but their contributions are less important in comparison to stronger transitions with f $\ge$ 0.01. For this reason many authors (and some codes) do not normally report the A- values for very weak transitions.

\section{Lifetimes}

The lifetime $\tau$ of a level $j$ is defined as follows:

\begin{equation}
{\tau}_j = \frac{1}{{\sum_{i}^{}} A_{ji}}.
\end{equation}

In Table 5 we include lifetimes for all 253 levels from our calculations from the {\sc grasp} code. These results {\em include} A- values from all types of transitions, i.e. E1, E2, M1 and M2. The only  available experimental result for a lifetime is for the 3s3p$^6$ $^2$S$_{1/2}$ level by Dumont {\em et al} \cite{dum}. Their value of 0.55$\pm$0.04 ns compares very well with our result (0.525 ns) and that of Bi{\'{e}}mont  and Tr{\" {a}}bert \cite{bt} (0.59 ns) and Berrington {\em et al}\,  \cite{bpw} (0.569 ns). However, other theoretical results for this level by Huang {\em et al} \cite{knh}, Fawcett \cite{bcf}, Dong {\em et al} \cite{dong} and Mohan {\em et al} \cite{mm1} vary between 0.35 and 0.95 ns -- see Table 5 of \cite{bpw}, and hence are comparatively less accurate. 

Since lifetimes for another 86 levels are available  from the calculations of Mohan {\em et al} \cite{mm1}, we compare these in Table 9 with our work with the {\sc grasp} code.  Two sets of $\tau$ are listed in Table 9, namely those with E1 contribution alone (GRASP3a) and those which also include the contributions from E2, M1 and  M2 (GRASP3b). For levels for which E1 transitions are possible,  contributions from the E2, M1 and  M2 transitions are negligible, but they are useful in determining $\tau$ for all levels -- see for example, levels 2/4/8. In the case of  levels of the  3s$^2$3p$^5$, 3s3p$^6$,  3s$^2$3p$^4$3d, 3s$^2$3p$^4$4s and 3s$^2$3p$^4$4p configurations, the $\tau$ of   Mohan {\em et al} differ by up to a factor of three for many -- see for example, 42--60, and in a majority of cases their results are higher. However, the discrepancy for the levels of the   3s$^2$3p$^4$4d configuration is greater, up to an order of magnitude for many, and two orders of magnitude for level 64, i.e.   3p$^4$($^3$P)4d  $^4$D$_{1/2}$. For all these levels the $\tau$ of  Mohan {\em et al} are invariably {\em higher}, because to span all  levels of the 3p$^4$4d configuration, a minimum of   253 levels (listed in Table 5) are required. Since  Mohan {\em et al} have not included the contribution of the missing 165 levels, their results for $\tau$ for the listed 88 levels are significantly overestimated.                                          

\section{Conclusions}

In this work, energy levels, radiative rates, oscillator strengths and line strengths for transitions among 253 fine-structure levels of Ti VI are computed  using the fully relativistic {\sc grasp} code, and results reported for electric and magnetic dipole and quadrupole transitions. For calculating these parameters an extensive CI (with up to 3749 levels) has been included, which has been observed to be very significant, particularly for the accurate determination of A- values and lifetimes. Furthermore, analogous calculations have been performed with the {\sc fac} code and with the inclusion of even larger CI with up to 18,459 levels, but the additional CI included does not appreciably affect the magnitude or orderings of the lowest 253 energy levels. Based on a variety of comparisons among different calculations, the reported energy levels are assessed to be accurate to better than 1\%. 

There is a paucity of measured energies for a majority of  the levels of Ti VI, and those  compiled by NIST are not as accurate as expected, because they are mostly based on earlier calculations involving limited CI. Additionally, the orderings of some of the levels are reversed in the NIST listings, and therefore a reassessment of their energy levels is highly desirable.

Earlier theoretical energies \cite{mm1} are available for up to 88 levels. Although the calculations of Mohan {\em et al} \cite{mm1} also included extensive CI, their energy levels are not as accurate as presented in this paper. Discrepancies are greater for the A- values between their data and the present calculations. As for the energy levels, extensive comparisons, based on a variety of calculations with the {\sc grasp} and {\sc fac} codes,  have been made for the A- values, and the accuracy of these is assessed to be  $\sim$ 20\% for a majority of the strong transitions. 

Lifetimes are also reported for all levels but  measurements are available for only one level of Ti VI for which there is no discrepancy with the present work.  However, the corresponding lifetimes of Mohan {\em et al} \cite{mm1} are significantly overestimated, by up to an order of magnitude, for a majority of the common 88 levels. This is mainly because they have not included the contribution of the levels of the 3s3p$^5$3d and 3s$^2$3p$^3$3d$^2$  configurations, which intermix with levels of the 3s$^2$3p$^5$, 3s3p$^6$,  3s$^2$3p$^4$3d,  3s3p$^5$3d, 3s$^2$3p$^3$3d$^2$,  3s$^2$3p$^4$4s, 3s$^2$3p$^4$4p and 3s$^2$3p$^4$4d  configurations, which they have included.

Finally, calculations for energy levels and radiative rates have been performed for up to 18,459 levels of Ti VI, but for brevity results have been reported for only the lowest 253 levels. However, a complete set of results for all calculated parameters can be obtained on request from one of the authors (K.Aggarwal@qub.ac.uk).

\section*{Acknowledgment}
Part of this work was performed at the Clark Atlanta University during a visit by KMA in June 2012 and he is thankful to the hospitality of CAU as well as to AWE Aldermaston for financial support. The work at CAU has been sponsored by DOE Office of Science and AFOSR, USA.
    



\clearpage

\begin{flushleft}
{\bf Table 1.} Target levels  of the $n$ = 3 configurations of Ti VI and their threshold energies (in Ryd).
\newline 
\end{flushleft}
{\small
\begin{tabular}{rllrrrrrrrrrr} \hline
  & & & & & & & & & \\
Index    & Configuration               & Level             &  NIST      & GRASP1     &   GRASP2     &  GRASP3  &  FAC1     & FAC2        &    HFR      &    MCHF    & CIV3       \\
  & & & & & & & & &  \\ \hline
  & & & & & & & & &  \\
   1      & 3s$^2$3p$^5$               &   $^2$P$^o_{3/2}$ &  0.00000	&  0.00000   &   0.00000    & 0.00000  &  0.00000  &  0.00000	 &  0.00000    &   0.00000   &  0.00000  \\
  2      & 3s$^2$3p$^5$     	       &   $^2$P$^o_{1/2}$ &  0.05312	&  0.05349   &   0.05253    & 0.05254  &  0.05190  &  0.05194	 &  0.05312    &   0.04614   &  0.05312  \\
  3      & 3s3p$^6$         	       &   $^2$S$_{1/2}$   &  1.79181	&  1.70921   &   1.80736    & 1.78017  &  1.78286  &  1.78395	 &  1.79176    &   1.77535   &  1.79133  \\
  4      & 3s$^2$3p$^4$ ($^3$P)3d      &   $^4$D$_{7/2}$   &            &  2.34891   &   2.50467    & 2.43742  &  2.41939  &  2.41741	 &  2.44068    &	     &  2.41651  \\
  5      & 3s$^2$3p$^4$ ($^3$P)3d      &   $^4$D$_{5/2}$   &            &  2.35198   &   2.50761    & 2.44032  &  2.42232  &  2.42034	 &  2.44308    &   2.42631   &  2.41925  \\
  6      & 3s$^2$3p$^4$ ($^3$P)3d      &   $^4$D$_{3/2}$   &            &  2.35665   &   2.51215    & 2.44487  &  2.42684  &  2.42486	 &  2.44724    &   2.42994   &  2.42359  \\
  7      & 3s$^2$3p$^4$ ($^3$P)3d      &   $^4$D$_{1/2}$   &            &  2.36066   &   2.51607    & 2.44881  &  2.43074  &  2.42877	 &  2.45105    &   2.43319   &  2.42740  \\
  8      & 3s$^2$3p$^4$ ($^3$P)3d      &   $^4$F$_{9/2}$   &            &  2.55963   &   2.73200    & 2.66259  &  2.64284  &  2.63970	 &  2.62930    &	     &  2.63250  \\
  9      & 3s$^2$3p$^4$ ($^1$D)3d      &   $^2$P$_{1/2}$   &  3.58715   &  2.61445   &   2.73481    & 2.67727  &  2.65491  &  2.64986	 &  2.65799    &   2.64166   &  3.67866  \\
 10      & 3s$^2$3p$^4$ ($^3$P)3d      &   $^4$F$_{7/2}$   &            &  2.57776   &   2.74964    & 2.68037  &  2.66045  &  2.65733	 &  2.64633    &	     &  2.65034  \\
 11      & 3s$^2$3p$^4$ ($^3$P)3d      &   $^4$F$_{5/2}$   &            &  2.59048   &   2.76228    & 2.69287  &  2.67287  &  2.66976	 &  2.68843    &   2.65636   &  2.66286  \\
 12      & 3s$^2$3p$^4$ ($^3$P)3d      &   $^4$F$_{3/2}$   &            &  2.59837   &   2.76649    & 2.70051  &  2.68048  &  2.67737	 &  2.66576    &   2.66283   &  2.67046  \\
 13      & 3s$^2$3p$^4$ ($^1$D)3d      &   $^2$P$_{3/2}$   &  3.56837   &  2.64550   &   2.77021    & 2.70884  &  2.68615  &  2.68111	 &  2.68768    &   2.66997   &  3.65922  \\
 14      & 3s$^2$3p$^4$ ($^3$P)3d      &   $^4$P$_{1/2}$   &  2.74671   &  2.68638   &   2.81951    & 2.77054  &  2.74822  &  2.74312	 &  2.72627    &   2.73440   &  2.73731  \\
 15      & 3s$^2$3p$^4$ ($^1$D)3d      &   $^2$D$_{3/2}$   &  3.68264   &  2.69733   &   2.83078    & 2.77465  &  2.75251  &  2.74841	 &  2.73098    &   2.73788   &  3.79308  \\
 16      & 3s$^2$3p$^4$ ($^3$P)3d      &   $^4$P$_{3/2}$   &            &  2.70483   &   2.84128    & 2.78345  &  2.76091  &  2.75596	 &  2.73988    &   2.74586   &  2.75149  \\
 17      & 3s$^2$3p$^4$ ($^3$P)3d      &   $^4$P$_{5/2}$   &            &  2.71212   &   2.84537    & 2.79540  &  2.77285  &  2.76787	 &  2.75126    &   2.75765   &  2.76331  \\
 18      & 3s$^2$3p$^4$ ($^1$D)3d      &   $^2$D$_{5/2}$   &  3.63805   &  2.73284   &   2.86894    & 2.80604  &  2.78320  &  2.77906	 &  2.76075    &   2.76541   &  3.74875  \\
 19      & 3s$^2$3p$^4$ ($^3$P)3d      &   $^2$F$_{7/2}$   &            &  2.74745   &   2.89542    & 2.83436  &  2.81121  &  2.80653	 &  2.77614    &	     &  2.79674  \\
 20      & 3s$^2$3p$^4$ ($^1$D)3d      &   $^2$G$_{9/2}$   &            &  2.77937   &   2.95585    & 2.87324  &  2.84997  &  2.84647	 &  2.81579    &	     &  2.83908  \\
 21      & 3s$^2$3p$^4$ ($^1$D)3d      &   $^2$G$_{7/2}$   &            &  2.78624   &   2.95562    & 2.87681  &  2.85348  &  2.84963	 &  2.81870    &	     &  2.84163  \\
 22      & 3s$^2$3p$^4$ ($^3$P)3d      &   $^2$F$_{5/2}$   &            &  2.79677   &   2.93693    & 2.87961  &  2.85591  &  2.85086	 &  2.81992    &   2.82840   &  2.84143  \\
 23      & 3s$^2$3p$^4$ ($^1$D)3d      &   $^2$F$_{5/2}$   &  3.01831   &  3.03169   &   3.16197    & 3.10321  &  3.07466  &  3.06810	 &  3.01817    &   3.04348   &  3.05835  \\
 24      & 3s$^2$3p$^4$ ($^1$D)3d      &   $^2$F$_{7/2}$   &            &  3.04294   &   3.17318    & 3.11455  &  3.08583  &  3.07927	 &  3.02926    &	     &  3.07002  \\
 25      & 3s$^2$3p$^4$ ($^1$S)3d      &   $^2$D$_{3/2}$   &  3.21336   &  3.27911   &   3.40778    & 3.28947  &  3.26967  &  3.26490	 &  3.24087    &   3.23478   &  3.25926  \\
 26      & 3s$^2$3p$^4$ ($^1$S)3d      &   $^2$D$_{5/2}$   &  3.22898   &  3.28966   &   3.41901    & 3.29928  &  3.27953  &  3.27489	 &  3.25170    &   3.24467   &  3.27015  \\
 27      & 3s$^2$3p$^4$ ($^1$D)3d      &   $^2$S$_{1/2}$   &  3.46167   &  3.70763   &   3.63458    & 3.54689  &  3.51941  &  3.51949	 &  3.46173    &   3.53341   &  3.58151  \\
 28      & 3s$^2$3p$^4$ ($^3$P)3d      &   $^2$P$_{3/2}$   &  2.65990	&  3.69560   &   3.88504    & 3.69319  &  3.66809  &  3.66011	 &  3.56857    &   3.62499   &  2.67311  \\
 29      & 3s$^2$3p$^4$ ($^3$P)3d      &   $^2$P$_{1/2}$   &  2.62820	&  3.73072   &   3.90475    & 3.71192  &  3.68665  &  3.67864	 &  3.58717    &   3.64205   &  2.64098  \\
 30      & 3s$^2$3p$^4$ ($^3$P)3d      &   $^2$D$_{5/2}$   &  2.75554	&  3.83147   &   3.93068    & 3.75087  &  3.72195  &  3.71724	 &  3.63742    &   3.70361   &  2.77788  \\
 31      & 3s$^2$3p$^4$ ($^3$P)3d      &   $^2$D$_{3/2}$   &  2.72461	&  3.87593   &   3.97650    & 3.79518  &  3.76592  &  3.76122	 &  3.68092    &   3.74244   &  2.74597  \\
  & & & & & & & & &  \\ \hline				  								   	   
\end{tabular}	
}												   		   
\begin {flushleft}														   
\begin{tabbing}
aaaaaaaaaaaaaaaaaaaaaaaaaaaaaaaaaaaa\= \kill
NIST:  {\tt  http://physics.nist.gov/PhysRefData/ASD/levels\_form.html} \\
GRASP1: Present results from 5  configurations and  60 levels \\
GRASP2: Present results from 16  configurations and  568 levels \\
GRASP3: Present results from 38  configurations and 3749 levels \\
FAC1: Present results with 5821 levels \\
FAC2: Present results with 9160 levels \\
HFR: Fawcett   \cite{bcf} \\ 
MCHF: Results of Forese-Fischer {\em et al} \cite{cff} \\ 
CIV3: Results of Mohan {\em et al} \cite{mm1}  \\ 
\end{tabbing}
\end {flushleft}
 
 \clearpage
 \begin{flushleft}
{\bf Table 2.} Levels  of the 3s$^2$3p$^4$4s  and 3s$^2$3p$^4$4p  configurations of Ti VI and their threshold energies (in Ryd).
\newline 
\end{flushleft}

\begin{tabular}{rllrrrrrrrrr} \hline
  & & & & & & & & & \\
Index    & Configuration               & Level                &  NIST      &  GRASP1   & GRASP3    &    FAC1  &  FAC2  &    MCHF   &    CIV3  \\ 
  & & & & & & & & &  \\ \hline
  & & & & & & & & &  \\
  1      & 3s$^2$3p$^4$ ($^3$P)4s      &   $^4$P$_{5/2}$      &  4.48458   & 4.44728   &  4.69652  & 4.66547  &  4.66185  &  4.47697 &  4.58739 \\
  2      & 3s$^2$3p$^4$ ($^3$P)4s      &   $^4$P$_{3/2}$      &  4.51142   & 4.47825   &  4.72714  & 4.69558  &  4.69192  &  4.50376 &  4.41655 \\
  3      & 3s$^2$3p$^4$ ($^3$P)4s      &   $^4$P$_{1/2}$      &  4.53254   & 4.49685   &  4.74492  & 4.71347  &  4.70981  &  4.52213 &  4.63504 \\
  4      & 3s$^2$3p$^4$ ($^3$P)4s      &   $^2$P$_{3/2}$      &  4.57976   & 4.55876   &  4.80687  & 4.78175  &  4.77779  &  4.57537 &  4.68360 \\
  5      & 3s$^2$3p$^4$ ($^3$P)4s      &   $^2$P$_{1/2}$      &  4.61495   & 4.59546   &  4.84274  & 4.81754  &  4.81355  &  4.60961 &  4.71846 \\
  6      & 3s$^2$3p$^4$ ($^1$D)4s      &   $^2$D$_{5/2}$      &  4.72763   & 4.72915   &  4.97540  & 4.94106  &  4.93714  &  4.72239 &  4.85198 \\
  7      & 3s$^2$3p$^4$ ($^1$D)4s      &   $^2$D$_{3/2}$      &  4.72869   & 4.73037   &  4.97680  & 4.94212  &  4.93818  &  4.72321 &  4.85305 \\
  8      & 3s$^2$3p$^4$ ($^1$S)4s      &   $^2$S$_{1/2}$      &  5.00281   & 5.06757   &  5.20215  & 5.16481  &  5.16104  &	     &  5.21634 \\
  9      & 3s$^2$3p$^4$ ($^3$P)4p      &   $^4$P$^o_{5/2}$    &  	   & 4.96996   &  5.21917  & 5.19009  &  5.18332  &	     &  4.95882 \\
 10      & 3s$^2$3p$^4$ ($^3$P)4p      &   $^4$P$^o_{3/2}$    &  	   & 4.97646   &  5.22433  & 5.19382  &  5.18941  &	     &  4.96585 \\
 11      & 3s$^2$3p$^4$ ($^3$P)4p      &   $^4$P$^o_{1/2}$    &  	   & 4.99161   &  5.23975  & 5.20812  &  5.20537  &	     &  5.07049 \\
 12      & 3s$^2$3p$^4$ ($^3$P)4p      &   $^4$D$^o_{7/2}$    &  	   & 5.02855   &  5.28874  & 5.25941  &  5.25725  &	     &  5.01609 \\
 13      & 3s$^2$3p$^4$ ($^3$P)4p      &   $^4$D$^o_{5/2}$    &  	   & 5.04188   &  5.30260  & 5.27339  &  5.27106  &	     &  5.02936 \\
 14      & 3s$^2$3p$^4$ ($^3$P)4p      &   $^4$D$^o_{3/2}$    &  	   & 5.05971   &  5.31891  & 5.28921  &  5.28681  &	     &  5.04712 \\
 15      & 3s$^2$3p$^4$ ($^3$P)4p      &   $^4$D$^o_{1/2}$    &  	   & 5.06670   &  5.32745  & 5.29715  &  5.29452  &	     &  5.05453 \\
 16      & 3s$^2$3p$^4$ ($^3$P)4p      &   $^2$D$^o_{5/2}$    &  	   & 5.08460   &  5.34638  & 5.31908  &  5.31650  &	     &  5.06974 \\
 17      & 3s$^2$3p$^4$ ($^3$P)4p      &   $^2$D$^o_{3/2}$    &  	   & 5.12013   &  5.37857  & 5.35117  &  5.36971  &	     &  5.10538 \\
 18      & 3s$^2$3p$^4$ ($^3$P)4p      &   $^2$P$^o_{1/2}$    &  	   & 5.08126   &  5.32409  & 5.29325  &  5.28987  &	     &  5.07049 \\
 19      & 3s$^2$3p$^4$ ($^3$P)4p      &   $^2$P$^o_{3/2}$    &  	   & 5.09599   &  5.34241  & 5.31111  &  5.30744  &	     &  5.08402 \\
 20      & 3s$^2$3p$^4$ ($^3$P)4p      &   $^4$S$^o_{3/2}$    &  	   & 5.14383   &  5.40337  & 5.37335  &  5.34834  &	     &  5.12919 \\
 21      & 3s$^2$3p$^4$ ($^3$P)4p      &   $^2$S$^o_{1/2}$    &  	   & 5.14853   &  5.40698  & 5.37842  &  5.37509  &	     &  5.13412 \\
 22      & 3s$^2$3p$^4$ ($^1$D)4p      &   $^2$F$^o_{5/2}$    &  	   & 5.25623   &  5.51086  & 5.47715  &  5.47078  &	     &  5.24663 \\
 23      & 3s$^2$3p$^4$ ($^1$D)4p      &   $^2$F$^o_{7/2}$    &  	   & 5.26595   &  5.52014  & 5.48548  &  5.48277  &	     &  5.25503 \\
 24      & 3s$^2$3p$^4$ ($^1$D)4p      &   $^2$P$^o_{3/2}$    &  	   & 5.41216   &  5.63655  & 5.62715  &  5.62098  &	     &  5.39876 \\
 25      & 3s$^2$3p$^4$ ($^1$D)4p      &   $^2$P$^o_{1/2}$    &  	   & 5.44063   &  5.66653  & 5.65486  &  5.64921  &	     &  5.42915 \\
 26      & 3s$^2$3p$^4$ ($^1$D)4p      &   $^2$D$^o_{3/2}$    &  	   & 5.33807   &  5.59800  & 5.56488  &  5.56158  &	     &  5.32692 \\
 27      & 3s$^2$3p$^4$ ($^1$D)4p      &   $^2$D$^o_{5/2}$    &  	   & 5.34404   &  5.60480  & 5.57096  &  5.47500  &	     &  5.33228 \\
 28      & 3s$^2$3p$^4$ ($^1$S)4p      &   $^2$P$^o_{3/2}$    &  	   & 5.64431   &  5.79149  & 5.75594  &  5.75338  &	     &  5.63645 \\
 29      & 3s$^2$3p$^4$ ($^1$S)4p      &   $^2$P$^o_{1/2}$    &  	   & 5.64060   &  5.78666  & 5.75596  &  5.75360  &	     &  5.63317 \\
  & & & & & & & & &  \\ \hline				  								   	   
\end{tabular}													   		   
\begin {flushleft}														   
\begin{tabbing}
aaaaaaaaaaaaaaaaaaaaaaaaaaaaaaaaaaaa\= \kill
NIST:  {\tt http://physics.nist.gov/PhysRefData/ASD/levels\_form.html} \\
GRASP1: Present results from 5  configurations and  60 levels \\
GRASP3: Present results from 38  configurations and 3749 levels \\
FAC1: Present results with 5821 levels \\
FAC2: Present results with 9160 levels \\
MCHF: Results of Forese-Fischer {\em et al} \cite{cff} \\ 
CIV3: Results of Mohan {\em et al}  \cite{mm1}  \\ 
\end{tabbing}
\end {flushleft}

\clearpage
\begin{flushleft}
{\bf Table 3.} Levels  of the 3s$^2$3p$^4$4d configuration of Ti VI and their threshold energies (in Ryd).
\newline 
\end{flushleft}

\begin{tabular}{rllrrrrrrrrr} \hline
  & & & & & & & & & \\
Index    & Configuration               & Level            &  NIST       &  GRASP1a &    GRASP3   &    FAC1   &   FAC2   &  CIV3  \\ 
  & & & & & & & & &  \\ \hline
  & & & & & & & & &  \\
  1      & 3s$^2$3p$^4$ ($^3$P)4d      &   $^4$D$_{7/2}$  &             & 5.82490  &  6.08962  & 6.06420  &  6.06454  &          \\   
  2      & 3s$^2$3p$^4$ ($^3$P)4d      &   $^4$D$_{5/2}$  &  	        & 5.82823  &  6.09278  & 6.06785  &  6.06818  &          \\   
  3      & 3s$^2$3p$^4$ ($^3$P)4d      &   $^4$D$_{3/2}$  &  	        & 5.83446  &  6.09890  & 6.07383  &  6.07415  &          \\   
  4      & 3s$^2$3p$^4$ ($^3$P)4d      &   $^4$D$_{1/2}$  &  	        & 5.84129  &  6.10573  & 6.07995  &  6.08028  &          \\   
  5      & 3s$^2$3p$^4$ ($^3$P)4d      &   $^4$F$_{9/2}$  &  	        & 5.86877  &  6.14363  & 6.11954  &  6.11999  &  6.13394 \\   
  6      & 3s$^2$3p$^4$ ($^3$P)4d      &   $^4$F$_{7/2}$  &  	        & 5.88180  &  6.15440  & 6.13085  &  6.13091  &  6.17280 \\   
  7      & 3s$^2$3p$^4$ ($^3$P)4d      &   $^4$F$_{5/2}$  &  5.95755    & 5.89798  &  6.16592  & 6.14114  &  6.14074  &  6.17841 \\   
  8      & 3s$^2$3p$^4$ ($^3$P)4d      &   $^4$P$_{1/2}$  &  	        & 5.90490  &  6.17516  & 6.15172  &  6.15216  &  6.16995 \\   
  9      & 3s$^2$3p$^4$ ($^3$P)4d      &   $^4$F$_{3/2}$  &  5.96427    & 5.90772  &  6.18903  & 6.14785  &  6.14738  &  6.18085 \\   
 10      & 3s$^2$3p$^4$ ($^3$P)4d      &   $^2$F$_{7/2}$  &  	        & 5.91105  &  6.18294  & 6.15829  &  6.15829  &  6.14426 \\   
 11      & 3s$^2$3p$^4$ ($^3$P)4d      &   $^4$P$_{3/2}$  &  	        & 5.91780  &  6.19444  & 6.16487  &  6.16529  &  6.18897 \\   
 12      & 3s$^2$3p$^4$ ($^3$P)4d      &   $^2$D$_{5/2}$  &  5.93467    & 5.92069  &  6.18697  & 6.16176  &  6.16174  &  6.15713 \\   
 13      & 3s$^2$3p$^4$ ($^3$P)4d      &   $^2$D$_{3/2}$  &  5.94110    & 5.93003  &  6.17325  & 6.16946  &  6.16956  &  6.16646 \\   
 14      & 3s$^2$3p$^4$ ($^3$P)4d      &   $^2$P$_{1/2}$  &  	        & 5.93522  &  6.20332  & 6.17835  &  6.17841  &  6.19575 \\   
 15      & 3s$^2$3p$^4$ ($^3$P)4d      &   $^4$P$_{5/2}$  &  	        & 5.93552  &  6.20583  & 6.18134  &  6.18144  &  6.20570 \\   
 16      & 3s$^2$3p$^4$ ($^3$P)4d      &   $^2$F$_{5/2}$  &  5.98189    & 5.94955  &  6.21391  & 6.18945  &  6.18920  &  6.19656 \\   
 17      & 3s$^2$3p$^4$ ($^3$P)4d      &   $^2$P$_{3/2}$  &  5.99923    & 5.97195  &  6.23767  & 6.21258  &  6.21243  &  6.22680 \\   
 18      & 3s$^2$3p$^4$ ($^1$D)4d      &   $^2$S$_{1/2}$  &  6.09300    & 6.09081  &  6.34510  & 6.31438  &  6.31444  &  6.35249 \\   
 19      & 3s$^2$3p$^4$ ($^1$D)4d      &   $^2$G$_{7/2}$  &  	        & 6.09261  &  6.36357  & 6.33367  &  6.33355  &  6.35075 \\   
 20      & 3s$^2$3p$^4$ ($^1$D)4d      &   $^2$G$_{9/2}$  &  	        & 6.09358  &  6.36490  & 6.33513  &  6.33502  &  6.35182 \\   
 21      & 3s$^2$3p$^4$ ($^1$D)4d      &   $^2$P$_{3/2}$  &  6.11548    & 6.11584  &  6.36876  & 6.33764  &  6.33715  &  6.37210 \\   
 22      & 3s$^2$3p$^4$ ($^1$D)4d      &   $^2$P$_{1/2}$  &  6.11960    & 6.12797  &  6.38025  & 6.34916  &  6.34851  &  6.38067 \\   
 23      & 3s$^2$3p$^4$ ($^1$D)4d      &   $^2$D$_{5/2}$  &  6.15294    & 6.14296  &  6.40333  & 6.37302  &  6.37279  &  6.40309 \\   
 24      & 3s$^2$3p$^4$ ($^1$D)4d      &   $^2$D$_{3/2}$  &  6.14465    & 6.15317  &  6.41307  & 6.38048  &  6.38238  &  6.39625 \\   
 25      & 3s$^2$3p$^4$ ($^1$D)4d      &   $^2$F$_{5/2}$  &  	        & 6.15703  &  6.43092  & 6.40220  &  6.40241  &  6.41100 \\   
 26      & 3s$^2$3p$^4$ ($^1$D)4d      &   $^2$F$_{7/2}$  &  	        & 6.15830  &  6.43256  & 6.40389  &  6.40407  &  6.41264 \\   
 27      & 3s$^2$3p$^4$ ($^1$S)4d      &   $^2$D$_{5/2}$  &  6.41790    & 6.46689  &  6.62817  & 6.59726  &  6.59737  &  6.63257 \\   
 28      & 3s$^2$3p$^4$ ($^1$S)4d      &   $^2$D$_{3/2}$  &  6.41778    & 6.46759  &  6.62852  & 6.59724  &  6.59735  &  6.63243 \\   

  & & & & & & & & &  \\ \hline				  								   	   
\end{tabular}													   		   
\begin {flushleft}														   
\begin{tabbing}
aaaaaaaaaaaaaaaaaaaaaaaaaaaaaaaaaaaa\= \kill
NIST  ({\tt http://physics.nist.gov/PhysRefData/ASD/levels\_form.html})\\
GRASP1a: Present results from 6  configurations and  88 levels \\
GRASP3: Present results from 38  configurations and 3749 levels \\
FAC1:  Present results with 5821 levels \\
FAC2:  Present results with 9160 levels \\
CIV3: Results of Mohan {\em et al} \cite{mm1}  \\ 
\end{tabbing}
\end {flushleft}

\clearpage
\begin{flushleft}
{\bf Table 4.} Levels  of the 3s$^2$3p$^4$4f configuration of Ti VI  and their threshold energies (in Ryd).
\newline 
\end{flushleft}

\begin{tabular}{rllrrrrrrrrr} \hline
  & & & & & & & & & \\
Index    & Configuration               & Level                 &  GRASP1b &    GRASP3	&    FAC1   &	FAC2  \\
  & & & & & & & & &  \\ \hline
  & & & & & & & & &  \\
  1      & 3s$^2$3p$^4$ ($^3$P)4f      &   $^4$F$^o_{9/2}$     & 6.35082  &  6.67048  &  6.63731 &  6.63571   \\    
  2      & 3s$^2$3p$^4$ ($^3$P)4f      &   $^4$F$^o_{7/2}$     & 6.35236  &  6.67099  &  6.63769 &  6.63609   \\    
  3      & 3s$^2$3p$^4$ ($^3$P)4f      &   $^4$F$^o_{5/2}$     & 6.35649  &  6.67382  &  6.63980 &  6.63827   \\    
  4      & 3s$^2$3p$^4$ ($^3$P)4f      &   $^4$F$^o_{3/2}$     & 6.36256  &  6.67954  &  6.64537 &  6.64379   \\    
  5      & 3s$^2$3p$^4$ ($^3$P)4f      &   $^4$G$^o_{11/2}$    & 6.39009  &  6.71824  &  6.68527 &  6.68382   \\    
  6      & 3s$^2$3p$^4$ ($^3$P)4f      &   $^2$F$^o_{7/2}$     & 6.39921  &  6.67619  &  6.64033 &  6.63911   \\    
  7      & 3s$^2$3p$^4$ ($^3$P)4f      &   $^2$G$^o_{9/2}$     & 6.39967  &  6.76323  &  6.70047 &  6.69893   \\    
  8      & 3s$^2$3p$^4$ ($^3$P)4f      &   $^2$F$^o_{5/2}$     & 6.40462  &  6.68227  &  6.64724 &  6.64550   \\    
  9      & 3s$^2$3p$^4$ ($^3$P)4f      &   $^4$D$^o_{1/2}$     & 6.40838  &  6.72650  &  6.69219 &  6.68967   \\    
 10      & 3s$^2$3p$^4$ ($^3$P)4f      &   $^4$D$^o_{3/2}$     & 6.41604  &  6.74044  &  6.66320 &  6.66193   \\    
 11      & 3s$^2$3p$^4$ ($^3$P)4f      &   $^4$G$^o_{9/2}$     & 6.42751  &  6.73472  &  6.73296 &  6.73119   \\    
 12      & 3s$^2$3p$^4$ ($^3$P)4f      &   $^4$G$^o_{7/2}$     & 6.43041  &  6.75295  &  6.71902 &  6.71753   \\    
 13      & 3s$^2$3p$^4$ ($^3$P)4f      &   $^4$D$^o_{5/2}$     & 6.43474  &  6.75425  &  6.67692 &  6.67253   \\    
 14      & 3s$^2$3p$^4$ ($^3$P)4f      &   $^4$G$^o_{5/2}$     & 6.43886  &  6.76210  &  6.71897 &  6.71749   \\    
 15      & 3s$^2$3p$^4$ ($^3$P)4f      &   $^2$D$^o_{3/2}$     & 6.44393  &  6.70016  &  6.70645 &  6.70253   \\    
 16      & 3s$^2$3p$^4$ ($^3$P)4f      &   $^4$D$^o_{7/2}$     & 6.44668  &  6.76542  &  6.73446 &  6.73245   \\    
 17      & 3s$^2$3p$^4$ ($^3$P)4f      &   $^2$G$^o_{7/2}$     & 6.45371  &  6.79739  &  6.76094 &  6.75916   \\    
 18      & 3s$^2$3p$^4$ ($^3$P)4f      &   $^2$D$^o_{5/2}$     & 6.47395  &  6.71316  &  6.72285 &  6.72221   \\    
 19      & 3s$^2$3p$^4$ ($^1$D)4f      &   $^2$P$^o_{1/2}$     & 6.60241  &  7.13759  &  6.74011 &  6.73644   \\    
 20      & 3s$^2$3p$^4$ ($^1$D)4f      &   $^2$P$^o_{3/2}$     & 6.60269  &  7.13874  &  6.74121 &  6.73782   \\    
 21      & 3s$^2$3p$^4$ ($^1$D)4f      &   $^2$H$^o_{11/2}$    & 6.61015  &  6.94584  &  6.90669 &  6.90494   \\    
 22      & 3s$^2$3p$^4$ ($^1$D)4f      &   $^2$H$^o_{9/2}$     & 6.61047  &  6.94506  &  6.90621 &  6.90394   \\    
 23      & 3s$^2$3p$^4$ ($^1$D)4f      &   $^2$D$^o_{5/2}$     & 6.63383  &  6.85289  &  6.81483 &  6.81274   \\    
 24      & 3s$^2$3p$^4$ ($^1$D)4f      &   $^2$D$^o_{3/2}$     & 6.64014  &  6.83641  &  6.79829 &  6.79608   \\    
 25      & 3s$^2$3p$^4$ ($^1$D)4f      &   $^2$F$^o_{7/2}$     & 6.65764  &  6.94752  &  6.90880 &  6.90611   \\    
 26      & 3s$^2$3p$^4$ ($^1$D)4f      &   $^2$F$^o_{5/2}$     & 6.66210  &  6.93817  &  6.89976 &  6.89698   \\    
 27      & 3s$^2$3p$^4$ ($^1$D)4f      &   $^2$G$^o_{7/2}$     & 6.66671  &  6.99981  &  6.96056 &  6.95831   \\    
 28      & 3s$^2$3p$^4$ ($^1$D)4f      &   $^2$G$^o_{9/2}$     & 6.66728  &  7.00337  &  6.96380 &  6.96152   \\ 
 29      & 3s$^2$3p$^4$ ($^1$S)4f      &   $^2$F$^o_{7/2}$     & 6.98197  &  7.18483  &  7.14229 &  7.14117   \\     
 30      & 3s$^2$3p$^4$ ($^1$S)4f      &   $^2$F$^o_{5/2}$     & 6.98316  &  7.18011  &  7.13842 &  7.13407   \\ 
  & & & & & & & & &  \\ \hline				  								   	   
\end{tabular}													   		   
\begin {flushleft}														   
\begin{tabbing}
aaaaaaaaaaaaaaaaaaaaaaaaaaaaaaaaaaaa\= \kill
GRASP1b: Present results from 7  configurations and  118 levels \\
GRASP3: Present results from 38  configurations and 3749 levels \\
FAC1:  Present results with 5821 levels \\
FAC2:  Present results with 9160 levels \\
\end{tabbing}
\end {flushleft}

\clearpage

\setcounter{table}{6}                                                                                         
\begin{table*}                                                                                                
\caption{Comparison of oscillator strengths for  transitions among the lowest 31 levels of  Ti VI. ($a{\pm}b \equiv$ $a\times$10$^{{\pm}b}$).}            
\begin{tabular}{rrlllllllll}                                                                                    
\hline                                                                                                        
\hline                                                                                                                                                                                                               
    $i$ & $j$    & 	 A (GRASP3)&     f (GRASP1)&    f (GRASP2) &    f (GRASP3) &     f (FAC1) &  f (FAC2)  &    f (MCHF)   &    f (CIV3)   &  R    \\				    
\hline                              	      	  					      						      
     1  &      3 &   1.3022$+$09   &	2.4157$-$2 &	3.3264$-$2 &  2.5578$-$2   &	2.523$-$2 & 2.508$-$2  &     2.581$-$2 &    2.060$-$2  & 0.7   \\
     1  &      5 &   1.3214$+$06   &	5.1087$-$5 &	3.4947$-$5 &  4.1435$-$5   &	3.978$-$5 & 4.024$-$5  &     2.392$-$5 &    3.013$-$5  & 1.3   \\
     1  &      6 &   8.7277$+$05   &	1.5708$-$5 &	1.4265$-$5 &  1.8178$-$5   &	1.691$-$5 & 1.690$-$5  &     1.260$-$5 &    1.477$-$5  & 0.6   \\ 
     1  &      7 &   3.2341$+$05   &	2.2003$-$6 &	2.5338$-$6 &  3.3571$-$6   &	3.770$-$6 & 3.635$-$6  &     2.662$-$6 &    2.741$-$6  & 0.3   \\
     1  &      9 &   3.5685$+$07   &	2.6142$-$4 &	1.6344$-$4 &  3.0990$-$4   &	2.973$-$4 & 2.825$-$4  &     2.672$-$4 &    2.241$-$4  & 0.2   \\
     1  &     11 &   6.4093$+$06   &	1.6798$-$4 &	1.3220$-$4 &  1.6505$-$4   &	1.686$-$4 & 1.691$-$4  &     1.104$-$4 &    1.311$-$4  & 1.0   \\
     1  &     12 &   5.8063$+$05   &	3.1549$-$5 &	1.8518$-$3 &  9.9119$-$6   &	5.069$-$6 & 3.884$-$7  &     7.934$-$6 &    3.996$-$7  & 1.7   \\
     1  &     13 &   1.8121$+$08   &	2.7064$-$3 &	6.8688$-$5 &  3.0745$-$3   &	2.954$-$3 & 2.838$-$3  &     2.450$-$3 &    2.324$-$3  & 0.3   \\
     1  &     14 &   4.3876$+$07   &	3.0295$-$4 &	3.3459$-$4 &  3.5581$-$4   &	3.606$-$4 & 3.613$-$4  &     2.520$-$4 &    2.870$-$4  & 1.1   \\
     1  &     15 &   8.7488$+$07   &	4.9097$-$4 &	2.4324$-$4 &  1.4148$-$3   &	1.357$-$3 & 1.355$-$3  &     1.104$-$3 &    1.018$-$3  & 0.8   \\
     1  &     16 &   2.1063$+$06   &	9.9916$-$4 &	9.1854$-$4 &  3.3845$-$5   &	2.695$-$5 & 4.880$-$5  &     1.071$-$5 &    2.495$-$4  & 0.1   \\
     1  &     17 &   2.7607$+$07   &	2.4504$-$4 &	1.5016$-$4 &  6.5973$-$4   &	5.835$-$4 & 5.126$-$4  &     4.455$-$4 &    2.806$-$4  & 0.8   \\
     1  &     18 &   1.3427$+$08   &	3.5888$-$3 &	2.9121$-$3 &  3.1843$-$3   &	2.983$-$3 & 3.153$-$3  &     2.633$-$3 &    3.023$-$3  & 0.7   \\
     1  &     22 &   3.5420$+$06   &	8.2587$-$5 &	3.7032$-$5 &  7.9768$-$5   &	6.810$-$5 & 6.725$-$5  &     5.276$-$5 &    6.677$-$5  & 0.8   \\
     1  &     23 &   3.6250$+$07   &	6.3453$-$4 &	5.8361$-$4 &  7.0295$-$4   &	7.158$-$4 & 7.210$-$4  &     4.749$-$4 &    5.675$-$4  & 1.0   \\
     1  &     25 &   8.4907$+$07   &	8.0809$-$4 &	5.7437$-$5 &  9.7687$-$4   &	8.816$-$4 & 8.076$-$4  &     6.673$-$4 &    5.677$-$4  & 0.8   \\
     1  &     26 &   8.9976$+$06   &	3.7004$-$5 &	1.6588$-$3 &  1.5436$-$4   &	7.989$-$5 & 1.699$-$5  &     3.075$-$5 &    5.217$-$5  & 0.6   \\
     1  &     27 &   8.5313$+$10   &	5.8702$-$1 &	4.1355$-$1 &  4.2212$-$1   &	4.169$-$1 & 4.215$-$1  &     4.577$-$1 &    4.652$-$1  & 1.1   \\
     1  &     28 &   1.1187$+$11   &	1.1790$-$0 &	1.0599$-$0 &  1.0211$-$0   &	1.017$-$0 & 1.020$-$0  &     1.014$-$0 &    9.944$-$1  & 0.9   \\
     1  &     29 &   3.3768$+$10   &	2.4254$-$1 &	1.6675$-$1 &  1.5255$-$1   &	1.527$-$1 & 1.521$-$1  &     1.397$-$1 &    1.285$-$1  & 0.9   \\
     1  &     30 &   1.3978$+$11   &	2.4577$-$0 &	1.8944$-$0 &  1.8554$-$0   &	1.829$-$0 & 1.842$-$0  &     1.910$-$0 &    1.878$-$0  & 1.0   \\
     1  &     31 &   9.9258$+$09   &	1.7962$-$1 &	7.2171$-$2 &  8.5792$-$2   &	8.079$-$2 & 8.450$-$2  &     1.124$-$1 &    1.108$-$1  & 1.0   \\
     2  &      3 &   6.0433$+$08   &	2.3909$-$2 &	3.2605$-$2 &  2.5207$-$2   &	2.491$-$2 & 2.477$-$2  &     2.665$-$2 &    2.066$-$2  & 0.7   \\
     2  &      6 &   1.1431$+$05   &	1.0501$-$5 &	4.5799$-$6 &  4.9731$-$6   &	4.803$-$6 & 5.162$-$6  &     2.162$-$6 &    2.620$-$6  & 2.8   \\
     2  &      7 &   5.5960$+$05   &	7.3214$-$6 &	9.0097$-$6 &  1.2133$-$5   &	1.169$-$5 & 1.167$-$5  &     9.259$-$6 &    1.069$-$5  & 0.3   \\
     2  &      9 &   1.3252$+$08   &	1.9734$-$3 &	1.5570$-$3 &  2.3948$-$3   &	2.316$-$3 & 2.210$-$3  &     1.918$-$3 &    1.773$-$3  & 0.3   \\
     2  &     12 &   8.1356$+$06   &	2.7412$-$4 &	1.4367$-$3 &  2.8890$-$4   &	3.246$-$4 & 3.727$-$4  &     1.988$-$4 &    3.258$-$4  & 1.0   \\
     2  &     13 &   5.6932$+$07   &	1.8906$-$3 &	1.2168$-$4 &  2.0090$-$3   &	1.896$-$3 & 1.780$-$3  &     1.564$-$3 &    1.418$-$3  & 0.4   \\
     2  &     14 &   6.0010$+$06   &	9.4429$-$5 &	1.0954$-$4 &  1.0113$-$4   &	9.586$-$5 & 8.947$-$5  &     7.407$-$5 &    7.121$-$5  & 0.9   \\
     2  &     15 &   1.0097$+$08   &	9.1736$-$4 &	2.2825$-$4 &  3.3927$-$3   &	3.074$-$3 & 3.029$-$3  &     2.723$-$3 &    1.915$-$3  & 0.7   \\
     2  &     16 &   1.4628$+$07   &	2.7189$-$3 &	2.8189$-$3 &  4.8838$-$4   &	4.637$-$4 & 6.104$-$4  &     2.797$-$4 &    1.332$-$3  & 0.9   \\
     2  &     25 &   9.0837$+$07   &	4.2760$-$3 &	7.7370$-$3 &  2.1586$-$3   &	2.521$-$3 & 3.010$-$3  &     2.070$-$3 &    3.771$-$3  & 1.1   \\
     2  &     27 &   3.2383$+$10   &	2.3707$-$1 &	3.5330$-$1 &  3.3017$-$1   &	3.290$-$1 & 3.289$-$1  &     3.192$-$1 &    3.035$-$1  & 1.1   \\
     2  &     28 &   6.7794$+$09   &	2.5081$-$1 &	1.0220$-$1 &  1.2735$-$1   &	1.205$-$1 & 1.266$-$1  &     1.713$-$1 &    1.674$-$1  & 0.9   \\
     2  &     29 &   8.3437$+$10   &	1.2313$-$0 &	7.7024$-$1 &  7.7570$-$1   &	7.670$-$1 & 7.740$-$1  &     8.183$-$1 &    8.186$-$1  & 0.9   \\
     2  &     31 &   1.2871$+$11   &	2.9086$-$0 &	2.3609$-$0 &  2.2879$-$0   &	2.261$-$0 & 2.270$-$0  &     2.312$-$0 &    2.273$-$0  & 1.0   \\
\hline                                                                                                        				      
\end{tabular} 
\begin{flushleft}
{\small
GRASP1: Present results from 3  configurations and  60 levels \\
GRASP2: Present results from 16  configurations and  568 levels \\
GRASP3: Present results from 38  configurations and 3749 levels \\
FAC1: Present results with 5821 levels \\
FAC2: Present results with 9160 levels \\
MCHF: Results of Forese-Fischer {\em et al} \cite{cff} \\ 
CIV3: Results of Mohan {\em et al} \cite{mm1} \\ 
R: Ratio of velocity and length forms of the f- values corresponding to GRASP3 calculations \\
}
\end{flushleft}

\end{table*}                                                                                                                                                                      
                                                                                                   

\setcounter{table}{7}                                                                                         
\begin{table*}                                                                                                
\caption{Comparison of oscillator strengths for some transitions of  Ti VI. ($a{\pm}b \equiv$ $a\times$10$^{{\pm}b}$).}            
\begin{tabular}{llllllllllll}                                                                                    
\hline                                                                                                        
\hline                                                                                                                                                                                                               
\multicolumn{2}{c}{Lower level} & \multicolumn{2}{c}{Upper level}    & 	   f (GRASP1a)&    f (GRASP3) &     f (FAC1) &  f (FAC2)  &     f (CIV3)       & R    \\	        		  
\hline                              	      	  					      
 3p$^5$  &  $^2$P$^o_{3/2}$   &  3p$^4$ ($^3$P)4s   &   $^4$P$_{5/2}$	&   7.020$-$4	  &  9.223$-$4    &  8.540$-$4  &      8.694$-$4  & 6.298$-$4  & 0.9  \\
 3p$^5$  &  $^2$P$^o_{3/2}$   &  3p$^4$ ($^3$P)4s   &   $^4$P$_{3/2}$	&   5.270$-$3	  &  6.246$-$3    &  4.830$-$3  &      4.882$-$3  & 6.283$-$3  & 0.9  \\
 3p$^5$  &  $^2$P$^o_{3/2}$   &  3p$^4$ ($^3$P)4s   &   $^4$P$_{1/2}$	&   7.713$-$7	  &  2.657$-$5    &  3.085$-$5  &      3.071$-$5  & 1.355$-$5  & 0.8  \\
 3p$^5$  &  $^2$P$^o_{3/2}$   &  3p$^4$ ($^3$P)4s   &   $^2$P$_{3/2}$	&   1.191$-$1	  &  1.348$-$1    &  1.283$-$1  &      1.289$-$1  & 1.129$-$0  & 0.9  \\
 3p$^5$  &  $^2$P$^o_{3/2}$   &  3p$^4$ ($^3$P)4s   &   $^2$P$_{1/2}$	&   2.560$-$2	  &  3.026$-$2    &  2.883$-$2  &      2.900$-$2  & 2.511$-$2  & 0.9  \\
 3p$^5$  &  $^2$P$^o_{3/2}$   &  3p$^4$ ($^1$D)4s   &   $^2$D$_{5/2}$	&   6.071$-$2	  &  7.784$-$2    &  7.340$-$2  &      7.407$-$2  & 7.120$-$2  & 0.9  \\
 3p$^5$  &  $^2$P$^o_{3/2}$   &  3p$^4$ ($^1$D)4s   &   $^2$D$_{3/2}$	&   1.308$-$3	  &  2.026$-$3    &  1.765$-$3  &      1.784$-$3  & 1.734$-$3  & 0.9  \\
 3p$^5$  &  $^2$P$^o_{3/2}$   &  3p$^4$ ($^1$S)4s   &   $^2$S$_{1/2}$	&   1.130$-$2	  &  1.539$-$2    &  1.552$-$2  &      1.578$-$2  & 1.481$-$2  & 0.8  \\
 3p$^5$  &  $^2$P$^o_{3/2}$   &  3p$^4$ ($^3$P)4d   &   $^4$D$_{5/2}$	&   3.702$-$6	  &  5.416$-$6    &  4.962$-$6  &      4.596$-$6  & 5.019$-$5  & 1.4  \\
 3p$^5$  &  $^2$P$^o_{3/2}$   &  3p$^4$ ($^3$P)4d   &   $^4$D$_{3/2}$	&   8.538$-$7	  &  1.277$-$5    &  1.303$-$5  &      1.298$-$5  & 7.723$-$5  & 0.4  \\
 3p$^5$  &  $^2$P$^o_{3/2}$   &  3p$^4$ ($^3$P)4d   &   $^4$D$_{1/2}$	&   6.630$-$8	  &  8.151$-$7    &  1.048$-$6  &      1.185$-$6  & 1.178$-$5  & 0.0  \\
 3p$^5$  &  $^2$P$^o_{3/2}$   &  3p$^4$ ($^3$P)4d   &   $^4$F$_{5/2}$	&   3.120$-$3	  &  1.158$-$2    &  1.386$-$2  &      1.434$-$2  & 1.280$-$2  & 0.8  \\
 3p$^5$  &  $^2$P$^o_{3/2}$   &  3p$^4$ ($^3$P)4d   &   $^4$P$_{1/2}$	&   1.241$-$4	  &  4.542$-$4    &  5.153$-$4  &      5.129$-$4  & 4.277$-$4  & 0.6  \\
 3p$^5$  &  $^2$P$^o_{3/2}$   &  3p$^4$ ($^3$P)4d   &   $^4$F$_{3/2}$	&   1.695$-$3	  &  1.075$-$4    &  4.667$-$3  &      4.710$-$3  & 1.438$-$3  & 1.0  \\
 3p$^5$  &  $^2$P$^o_{3/2}$   &  3p$^4$ ($^3$P)4d   &   $^4$P$_{3/2}$	&   4.740$-$4	  &  9.237$-$4    &  1.905$-$4  &      1.920$-$4  & 1.708$-$3  & 0.7  \\
 3p$^5$  &  $^2$P$^o_{3/2}$   &  3p$^4$ ($^3$P)4d   &   $^2$D$_{5/2}$	&   5.318$-$3	  &  6.846$-$3    &  8.047$-$3  &      7.498$-$3  & 1.566$-$2  & 0.8  \\
 3p$^5$  &  $^2$P$^o_{3/2}$   &  3p$^4$ ($^3$P)4d   &   $^2$D$_{3/2}$	&   1.742$-$3	  &  4.010$-$3    &  8.485$-$4  &      6.930$-$4  & 6.361$-$3  & 0.9  \\
 3p$^5$  &  $^2$P$^o_{3/2}$   &  3p$^4$ ($^3$P)4d   &   $^2$P$_{1/2}$	&   2.695$-$4	  &  8.596$-$4    &  9.558$-$4  &      9.305$-$4  & 1.474$-$3  & 0.6  \\
 3p$^5$  &  $^2$P$^o_{3/2}$   &  3p$^4$ ($^3$P)4d   &   $^4$P$_{5/2}$	&   2.939$-$5	  &  2.046$-$5    &  6.673$-$5  &      1.743$-$4  & 1.066$-$2  & 0.7  \\
 3p$^5$  &  $^2$P$^o_{3/2}$   &  3p$^4$ ($^3$P)4d   &   $^2$F$_{5/2}$	&   7.549$-$3	  &  7.725$-$3    &  7.986$-$3  &      7.504$-$3  & 3.082$-$3  & 0.8  \\
 3p$^5$  &  $^2$P$^o_{3/2}$   &  3p$^4$ ($^3$P)4d   &   $^2$P$_{3/2}$	&   7.505$-$4	  &  2.119$-$3    &  2.466$-$3  &      2.393$-$3  & 4.562$-$3  & 0.6  \\
 3p$^5$  &  $^2$P$^o_{3/2}$   &  3p$^4$ ($^1$D)4d   &   $^2$S$_{1/2}$	&   1.128$-$3	  &  8.542$-$3    &  9.308$-$3  &      9.289$-$3  & 6.757$-$3  & 0.5  \\
 3p$^5$  &  $^2$P$^o_{3/2}$   &  3p$^4$ ($^1$D)4d   &   $^2$P$_{3/2}$	&   1.558$-$2	  &  8.452$-$3    &  9.953$-$3  &      1.012$-$2  & 1.739$-$2  & 1.2  \\
 3p$^5$  &  $^2$P$^o_{3/2}$   &  3p$^4$ ($^1$D)4d   &   $^2$P$_{1/2}$	&   2.990$-$3	  &  1.239$-$3    &  1.514$-$3  &      1.522$-$3  & 3.307$-$3  & 1.5  \\
 3p$^5$  &  $^2$P$^o_{3/2}$   &  3p$^4$ ($^1$D)4d   &   $^2$D$_{5/2}$	&   5.517$-$3	  &  1.245$-$2    &  1.414$-$2  &      1.405$-$2  & 1.546$-$2  & 0.7  \\
 3p$^5$  &  $^2$P$^o_{3/2}$   &  3p$^4$ ($^1$D)4d   &   $^2$D$_{3/2}$	&   9.246$-$4	  &  1.931$-$3    &  2.201$-$3  &      2.156$-$3  & 2.806$-$3  & 0.7  \\
 3p$^5$  &  $^2$P$^o_{3/2}$   &  3p$^4$ ($^1$D)4d   &   $^2$F$_{5/2}$	&   5.513$-$4	  &  3.734$-$4    &  4.028$-$4  &      3.892$-$4  & 9.190$-$4  & 0.7  \\
 3p$^5$  &  $^2$P$^o_{3/2}$   &  3p$^4$ ($^1$S)4d   &   $^2$D$_{5/2}$	&   1.852$-$3	  &  1.655$-$3    &  1.649$-$3  &      1.748$-$3  & 3.502$-$3  & 0.8  \\
 3p$^5$  &  $^2$P$^o_{3/2}$   &  3p$^4$ ($^1$S)4d   &   $^2$D$_{3/2}$	&   1.788$-$4	  &  2.244$-$4    &  2.244$-$4  &      2.454$-$4  & 3.586$-$4  & 0.8  \\
 \hline                                                                                                         					
\end{tabular} 																		
 \end{table*}   																	
 																			
 \setcounter{table}{7}                                                                                          					
\begin{table*}                                                                                                						
\caption{Comparison of oscillator strengths for some transitions of  Ti VI. ($a{\pm}b \equiv$ $a\times$10$^{{\pm}b}$).}            			
\begin{tabular}{llllllllllll}                                                                                    					
\hline                                                                                                        						
\hline                                                                                                                                                          						   
\multicolumn{2}{c}{Lower level} & \multicolumn{2}{c}{Upper level}    & 	   f (GRASP1a)&    f (GRASP3) &     f (FAC1) &  f (FAC2)  &     f (CIV3)       & R    \\	        		  
\hline     																		
 3p$^5$  &  $^2$P$^o_{1/2}$   &  3p$^4$ ($^3$P)4s   &   $^4$P$_{3/2}$	&   6.290$-$4	  &  6.767$-$4    &  4.654$-$4  &      4.714$-$4  & 7.937$-$4  & 0.9  \\
 3p$^5$  &  $^2$P$^o_{1/2}$   &  3p$^4$ ($^3$P)4s   &   $^4$P$_{1/2}$	&   1.054$-$3	  &  1.336$-$3    &  1.333$-$3  &      1.352$-$3  & 9.324$-$4  & 0.9  \\
 3p$^5$  &  $^2$P$^o_{1/2}$   &  3p$^4$ ($^3$P)4s   &   $^2$P$_{3/2}$	&   3.209$-$2	  &  3.530$-$2    &  3.280$-$2  &      3.295$-$2  & 2.923$-$2  & 0.9  \\
 3p$^5$  &  $^2$P$^o_{1/2}$   &  3p$^4$ ($^3$P)4s   &   $^2$P$_{1/2}$	&   9.008$-$2	  &  9.828$-$2    &  9.206$-$2  &      9.243$-$2  & 8.654$-$2  & 0.9  \\
 3p$^5$  &  $^2$P$^o_{1/2}$   &  3p$^4$ ($^1$D)4s   &   $^2$D$_{3/2}$	&   8.629$-$2	  &  1.091$-$1    &  1.035$-$1  &      1.042$-$1  & 9.714$-$2  & 0.9  \\
 3p$^5$  &  $^2$P$^o_{1/2}$   &  3p$^4$ ($^1$S)4s   &   $^2$S$_{1/2}$	&   1.870$-$2	  &  2.725$-$2    &  2.760$-$2  &      2.803$-$2  & 2.159$-$2  & 0.9  \\
 3p$^5$  &  $^2$P$^o_{1/2}$   &  3p$^4$ ($^3$P)4d   &   $^4$D$_{3/2}$	&   5.126$-$5	  &  6.401$-$5    &  6.824$-$5  &      6.751$-$5  & 1.911$-$4  & 1.2  \\
 3p$^5$  &  $^2$P$^o_{1/2}$   &  3p$^4$ ($^3$P)4d   &   $^4$D$_{1/2}$	&   3.888$-$7	  &  1.354$-$5    &  1.200$-$5  &      1.034$-$5  & 6.384$-$5  & 0.2  \\
 3p$^5$  &  $^2$P$^o_{1/2}$   &  3p$^4$ ($^3$P)4d   &   $^4$P$_{1/2}$	&   4.409$-$5	  &  2.589$-$6    &  2.676$-$6  &      2.003$-$6  & 1.108$-$4  & 1.6  \\
 3p$^5$  &  $^2$P$^o_{1/2}$   &  3p$^4$ ($^3$P)4d   &   $^4$F$_{3/2}$	&   2.879$-$3	  &  4.162$-$5    &  1.301$-$2  &      1.337$-$2  & 1.167$-$3  & 1.0  \\
 3p$^5$  &  $^2$P$^o_{1/2}$   &  3p$^4$ ($^3$P)4d   &   $^4$P$_{3/2}$	&   2.620$-$5	  &  5.748$-$3    &		&		  & 9.901$-$3  & 0.8  \\
 3p$^5$  &  $^2$P$^o_{1/2}$   &  3p$^4$ ($^3$P)4d   &   $^2$D$_{3/2}$	&   6.452$-$3	  &  1.084$-$2    &  5.986$-$3  &      5.505$-$3  & 1.401$-$2  & 0.8  \\
 3p$^5$  &  $^2$P$^o_{1/2}$   &  3p$^4$ ($^3$P)4d   &   $^2$P$_{1/2}$	&   7.741$-$4	  &  1.468$-$3    &  1.623$-$3  &      1.534$-$3  & 3.836$-$3  & 0.6  \\
 3p$^5$  &  $^2$P$^o_{1/2}$   &  3p$^4$ ($^3$P)4d   &   $^2$P$_{3/2}$	&   4.565$-$3	  &  6.213$-$3    &  6.833$-$3  &      6.624$-$3  & 1.482$-$2  & 0.7  \\
 3p$^5$  &  $^2$P$^o_{1/2}$   &  3p$^4$ ($^1$D)4d   &   $^2$S$_{1/2}$	&   4.573$-$4	  &  7.438$-$3    &  7.965$-$3  &      7.911$-$3  & 5.533$-$3  & 0.4  \\
 3p$^5$  &  $^2$P$^o_{1/2}$   &  3p$^4$ ($^1$D)4d   &   $^2$P$_{3/2}$	&   7.899$-$3	  &  4.674$-$3    &  5.453$-$3  &      5.570$-$3  & 9.329$-$3  & 1.2  \\
 3p$^5$  &  $^2$P$^o_{1/2}$   &  3p$^4$ ($^1$D)4d   &   $^2$P$_{1/2}$	&   1.560$-$2	  &  1.109$-$2    &  1.294$-$2  &      1.307$-$2  & 2.098$-$2  & 1.1  \\
 3p$^5$  &  $^2$P$^o_{1/2}$   &  3p$^4$ ($^1$D)4d   &   $^2$D$_{3/2}$	&   9.666$-$3	  &  2.001$-$2    &  2.282$-$2  &      2.262$-$2  & 2.645$-$2  & 0.7  \\
 3p$^5$  &  $^2$P$^o_{1/2}$   &  3p$^4$ ($^1$S)4d   &   $^2$D$_{3/2}$	&   3.204$-$3	  &  3.530$-$3    &  3.735$-$3  &      3.840$-$3  & 6.157$-$3  & 0.8  \\
\hline                                                                                                        
\end{tabular} 
\begin{flushleft}
{\small
GRASP1a: Present results from 6  configurations and  88 levels \\
GRASP3: Present results from 38  configurations and 3749 levels \\
FAC1: Present results with 5821 levels \\
FAC2: Present results with 9160 levels \\
CIV3: Results of Mohan {\em et al} \cite{mm1}  \\ 
R: Ratio of velocity and length forms of the f- values corresponding to GRASP3 calculations \\
}
\end{flushleft}

\end{table*}                                                                                                                                                                      
                                                                                                   

\clearpage
\begin{flushleft}
{\bf Table 9.} Lifetimes (s) of some levels of Ti VI. ($a{\pm}b \equiv$ $a\times$10$^{{\pm}b}$).
\newline 
\end{flushleft}

\begin{tabular}{rllrrrrrrrrr} \hline
  & & & & & & & & & \\
Index    & Configuration              & Level               &  GRASP3a   &  GRASP3b        & CIV3          \\
  & & & & & & & & &  \\ \hline
  & & & & & & & & &  \\                                                                             $ $
   1	& 3s$^2$3p$^5$  	      &   $^2$P$^o_{3/2}$    &            &                  &      $ $    \\
   2	& 3s$^2$3p$^5$  	      &   $^2$P$^o_{1/2}$    &            &    2.902$-$01    &	    $ $    \\
   3	& 3s3p$^6$		      &   $^2$S$_{1/2}$      & 5.245$-$10 &    5.245$-$10    &  6.40$-$10  \\
   4	& 3s$^2$3p$^4$ ($^3$P)3d      &   $^4$D$_{7/2}$      &            &    5.840$-$02    &	    $ $    \\
   5	& 3s$^2$3p$^4$ ($^3$P)3d      &   $^4$D$_{5/2}$      & 7.568$-$07 &    7.568$-$07    &  1.06$-$06  \\
   6	& 3s$^2$3p$^4$ ($^3$P)3d      &   $^4$D$_{3/2}$      & 1.013$-$06 &    1.013$-$06    &  1.32$-$06  \\
   7	& 3s$^2$3p$^4$ ($^3$P)3d      &   $^4$D$_{1/2}$      & 1.132$-$06 &    1.132$-$06    &  1.35$-$06  \\
   8	& 3s$^2$3p$^4$ ($^3$P)3d      &   $^4$F$_{9/2}$      &            &    6.398$-$01    &	    $ $    \\
   9	& 3s$^2$3p$^4$ ($^1$D)3d      &   $^2$P$_{1/2}$      & 5.945$-$09 &    5.945$-$09    &  8.30$-$09  \\
  10	& 3s$^2$3p$^4$ ($^3$P)3d      &   $^4$F$_{7/2}$      &            &    6.147$-$01    &	    $ $    \\
  11	& 3s$^2$3p$^4$ ($^3$P)3d      &   $^4$F$_{5/2}$      & 1.560$-$07 &    1.560$-$07    &  2.01$-$07  \\
  12	& 3s$^2$3p$^4$ ($^3$P)3d      &   $^4$F$_{3/2}$      & 1.147$-$07 &    1.147$-$07    &  1.11$-$07  \\
  13	& 3s$^2$3p$^4$ ($^1$D)3d      &   $^2$P$_{3/2}$      & 4.199$-$09 &    4.199$-$09    &  5.80$-$09  \\
  14	& 3s$^2$3p$^4$ ($^3$P)3d      &   $^4$P$_{1/2}$      & 2.005$-$08 &    2.005$-$08    &  2.59$-$08  \\
  15	& 3s$^2$3p$^4$ ($^1$D)3d      &   $^2$D$_{3/2}$      & 5.306$-$09 &    5.306$-$09    &  8.52$-$09  \\
  16	& 3s$^2$3p$^4$ ($^3$P)3d      &   $^4$P$_{3/2}$      & 5.976$-$08 &    5.976$-$08    &  1.85$-$08  \\
  17	& 3s$^2$3p$^4$ ($^3$P)3d      &   $^4$P$_{5/2}$      & 3.622$-$08 &    3.622$-$08    &  8.72$-$08  \\
  18	& 3s$^2$3p$^4$ ($^1$D)3d      &   $^2$D$_{5/2}$      & 7.448$-$09 &    7.448$-$09    &  8.01$-$09  \\
  19	& 3s$^2$3p$^4$ ($^3$P)3d      &   $^2$F$_{7/2}$      &            &    7.814$-$02    &	    $ $    \\
  20	& 3s$^2$3p$^4$ ($^1$D)3d      &   $^2$G$_{9/2}$      &            &    1.377$-$01    &	    $ $    \\
  21	& 3s$^2$3p$^4$ ($^1$D)3d      &   $^2$G$_{7/2}$      &            &    1.281$-$01    &	    $ $    \\
  22	& 3s$^2$3p$^4$ ($^3$P)3d      &   $^2$F$_{5/2}$      & 2.823$-$07 &    2.823$-$07    &  3.46$-$07  \\
  23	& 3s$^2$3p$^4$ ($^1$D)3d      &   $^2$F$_{5/2}$      & 2.759$-$08 &    2.759$-$08    &  3.52$-$08  \\
  24	& 3s$^2$3p$^4$ ($^1$D)3d      &   $^2$F$_{7/2}$      &            &    4.278$-$02    &	    $ $    \\
  25	& 3s$^2$3p$^4$ ($^1$S)3d      &   $^2$D$_{3/2}$      & 5.690$-$09 &    5.690$-$09    &  4.90$-$09  \\
  26	& 3s$^2$3p$^4$ ($^1$S)3d      &   $^2$D$_{5/2}$      & 1.111$-$07 &    1.111$-$07    &  3.35$-$07  \\
  27	& 3s$^2$3p$^4$ ($^1$D)3d      &   $^2$S$_{1/2}$      & 8.496$-$12 &    8.496$-$12    &  7.92$-$12  \\
  28	& 3s$^2$3p$^4$ ($^3$P)3d      &   $^2$P$_{3/2}$      & 8.428$-$12 &    8.428$-$12    &  8.64$-$12  \\
  29	& 3s$^2$3p$^4$ ($^3$P)3d      &   $^2$P$_{1/2}$      & 8.532$-$12 &    8.532$-$12    &  8.75$-$12  \\
  30	& 3s$^2$3p$^4$ ($^3$P)3d      &   $^2$D$_{5/2}$      & 7.154$-$12 &    7.154$-$12    &  7.07$-$12  \\
  31	& 3s$^2$3p$^4$ ($^3$P)3d      &   $^2$D$_{3/2}$      & 7.213$-$12 &    7.213$-$12    &  7.12$-$12  \\
  32	& 3s$^2$3p$^4$ ($^3$P)4s      &   $^4$P$_{5/2}$      & 9.170$-$09 &    9.157$-$09    &  1.41$-$10  \\
  33	& 3s$^2$3p$^4$ ($^3$P)4s      &   $^4$P$_{3/2}$      & 8.469$-$10 &    8.468$-$10    &  8.76$-$10  \\
  34	& 3s$^2$3p$^4$ ($^3$P)4s      &   $^4$P$_{1/2}$      & 4.064$-$09 &    4.062$-$09    &  6.18$-$09  \\
  35	& 3s$^2$3p$^4$ ($^3$P)4s      &   $^2$P$_{3/2}$      & 3.542$-$11 &    3.542$-$11    &  4.46$-$11  \\
  36	& 3s$^2$3p$^4$ ($^3$P)4s      &   $^2$P$_{1/2}$      & 3.388$-$11 &    3.388$-$11    &  4.15$-$11  \\
  37	& 3s$^2$3p$^4$ ($^1$D)4s      &   $^2$D$_{5/2}$      & 9.692$-$11 &    9.692$-$11    &  1.11$-$10  \\
  38	& 3s$^2$3p$^4$ ($^1$D)4s      &   $^2$D$_{3/2}$      & 9.067$-$11 &    9.067$-$11    &  1.07$-$10  \\
  39	& 3s$^2$3p$^4$ ($^1$S)4s      &   $^2$S$_{1/2}$      & 8.002$-$11 &    8.002$-$11    &  9.01$-$11  \\
   & & & & & & & & &  \\ \hline				  								   	   
\end{tabular} 

\clearpage
\begin{flushleft}
{\bf Table 9.} Lifetimes (s) of some levels of Ti VI. ($a{\pm}b \equiv$ $a\times$10$^{{\pm}b}$).
\newline 
\end{flushleft}

\begin{tabular}{rllrrrrrrrrr} \hline
  & & & & & & & & & \\
Index    & Configuration              & Level               &  GRASP3a   &  GRASP3b        & CIV3          \\
  & & & & & & & & &  \\ \hline
  & & & & & & & & &  \\     
  40	& 3s$^2$3p$^4$ ($^3$P)4p      &   $^4$P$^o_{5/2}$    & 9.389$-$11 &    9.389$-$11    &  1.76$-$10  \\
  41	& 3s$^2$3p$^4$ ($^3$P)4p      &   $^4$P$^o_{3/2}$    & 1.136$-$10 &    1.136$-$10    &  1.76$-$10  \\
  42	& 3s$^2$3p$^4$ ($^3$P)4p      &   $^4$P$^o_{1/2}$    & 8.892$-$11 &    8.892$-$11    &  1.77$-$10  \\
  43	& 3s$^2$3p$^4$ ($^3$P)4p      &   $^4$D$^o_{7/2}$    & 8.772$-$11 &    8.772$-$11    &  1.72$-$10  \\
  44	& 3s$^2$3p$^4$ ($^3$P)4p      &   $^4$D$^o_{5/2}$    & 9.087$-$11 &    9.087$-$11    &  1.73$-$10  \\
  45	& 3s$^2$3p$^4$ ($^3$P)4p      &   $^4$D$^o_{3/2}$    & 8.572$-$11 &    8.572$-$11    &  1.74$-$10  \\
  46	& 3s$^2$3p$^4$ ($^3$P)4p      &   $^4$D$^o_{1/2}$    & 9.448$-$11 &    9.448$-$11    &  5.42$-$10  \\  
  47	& 3s$^2$3p$^4$ ($^3$P)4p      &   $^2$D$^o_{5/2}$    & 1.067$-$10  &  1.067$-$10    & 2.51$-$10   \\
  48	& 3s$^2$3p$^4$ ($^3$P)4p      &   $^2$D$^o_{3/2}$    & 1.083$-$10  &  1.083$-$10    & 2.49$-$10   \\
  49	& 3s$^2$3p$^4$ ($^3$P)4p      &   $^2$P$^o_{1/2}$    & 9.348$-$11  &  9.348$-$11    & 2.75$-$10   \\
  50	& 3s$^2$3p$^4$ ($^3$P)4p      &   $^2$P$^o_{3/2}$    & 1.048$-$10  &  1.048$-$10    & 2.80$-$10   \\
  51	& 3s$^2$3p$^4$ ($^3$P)4p      &   $^4$S$^o_{3/2}$    & 7.393$-$11  &  7.393$-$11    & 1.62$-$10   \\
  52	& 3s$^2$3p$^4$ ($^3$P)4p      &   $^2$S$^o_{1/2}$    & 1.183$-$10  &  1.183$-$10    & 2.99$-$10   \\
  53	& 3s$^2$3p$^4$ ($^1$D)4p      &   $^2$F$^o_{5/2}$    & 8.879$-$11  &  8.878$-$11    & 1.96$-$10   \\
  54	& 3s$^2$3p$^4$ ($^1$D)4p      &   $^2$F$^o_{7/2}$    & 8.867$-$11  &  8.866$-$11    & 1.92$-$10   \\
  55	& 3s$^2$3p$^4$ ($^1$D)4p      &   $^2$P$^o_{3/2}$    & 9.499$-$11  &  9.499$-$11    & 2.07$-$10   \\
  56	& 3s$^2$3p$^4$ ($^1$D)4p      &   $^2$P$^o_{1/2}$    & 9.355$-$11  &  9.355$-$11    & 2.05$-$10   \\
  57	& 3s$^2$3p$^4$ ($^1$D)4p      &   $^2$D$^o_{3/2}$    & 8.520$-$11  &  8.519$-$11    & 1.95$-$10   \\
  58	& 3s$^2$3p$^4$ ($^1$D)4p      &   $^2$D$^o_{5/2}$    & 8.282$-$11  &  8.282$-$11    & 1.94$-$10   \\
  59	& 3s$^2$3p$^4$ ($^1$S)4p      &   $^2$P$^o_{3/2}$    & 8.284$-$11  &  8.284$-$11    & 1.97$-$10   \\
  60	& 3s$^2$3p$^4$ ($^1$S)4p      &   $^2$P$^o_{1/2}$    & 8.358$-$11  &  8.358$-$11    & 1.98$-$10   \\
   & & & & & & & & &  \\ \hline				  								   	   
\end{tabular} 

\clearpage
\begin{flushleft}
{\bf Table 9.} Lifetimes (s) of some levels of Ti VI. ($a{\pm}b \equiv$ $a\times$10$^{{\pm}b}$).
\newline 
\end{flushleft}

\begin{tabular}{rllrrrrrrrrr} \hline
  & & & & & & & & & \\
Index    & Configuration              & Level               &  GRASP3a   &  GRASP3b        & CIV3          \\
  & & & & & & & & &  \\ \hline
  & & & & & & & & &  \\     
  61	& 3s$^2$3p$^4$ ($^3$P)4d      &   $^4$D$_{7/2}$      & 3.089$-$10  &  3.088$-$10    &	  $ $	  \\
  62	& 3s$^2$3p$^4$ ($^3$P)4d      &   $^4$D$_{5/2}$      & 3.050$-$10  &  3.050$-$10    & 9.99$-$08   \\
  63	& 3s$^2$3p$^4$ ($^3$P)4d      &   $^4$D$_{3/2}$      & 3.027$-$10  &  3.026$-$10    & 1.95$-$08   \\
  64	& 3s$^2$3p$^4$ ($^3$P)4d      &   $^4$D$_{1/2}$      & 3.039$-$10  &  3.039$-$10    & 3.86$-$08   \\
  65	& 3s$^2$3p$^4$ ($^3$P)4d      &   $^4$F$_{9/2}$      & 3.022$-$10  &  3.021$-$10    &	  $ $	  \\
  66	& 3s$^2$3p$^4$ ($^3$P)4d      &   $^4$F$_{7/2}$      & 3.053$-$10  &  3.053$-$10    &	  $ $	  \\
  67	& 3s$^2$3p$^4$ ($^3$P)4d      &   $^4$F$_{5/2}$      & 1.878$-$10  &  1.878$-$10    & 3.82$-$10   \\
  68	& 3s$^2$3p$^4$ ($^3$P)4d      &   $^4$P$_{1/2}$      & 2.617$-$10  &  2.617$-$10    & 3.39$-$09   \\
  69	& 3s$^2$3p$^4$ ($^3$P)4d      &   $^4$F$_{3/2}$      & 2.884$-$10  &  2.884$-$10    & 1.62$-$09   \\
  70	& 3s$^2$3p$^4$ ($^3$P)4d      &   $^2$F$_{7/2}$      & 3.127$-$10  &  3.127$-$10    &	  $ $	  \\
  71	& 3s$^2$3p$^4$ ($^3$P)4d      &   $^4$P$_{3/2}$      & 2.270$-$10  &  2.270$-$10    & 4.94$-$10   \\
  72	& 3s$^2$3p$^4$ ($^3$P)4d      &   $^2$D$_{5/2}$      & 2.205$-$10  &  2.204$-$10    & 3.14$-$09   \\
  73	& 3s$^2$3p$^4$ ($^3$P)4d      &   $^2$D$_{3/2}$      & 1.733$-$10  &  1.733$-$10    & 2.47$-$09   \\
  74	& 3s$^2$3p$^4$ ($^3$P)4d      &   $^2$P$_{1/2}$      & 2.411$-$10  &  2.411$-$10    & 4.83$-$10   \\
  75	& 3s$^2$3p$^4$ ($^3$P)4d      &   $^4$P$_{5/2}$      & 2.972$-$10  &  2.971$-$10    & 4.55$-$10   \\
  76	& 3s$^2$3p$^4$ ($^3$P)4d      &   $^2$F$_{5/2}$      & 2.150$-$10  &  2.150$-$10    & 1.58$-$09   \\
  77	& 3s$^2$3p$^4$ ($^3$P)4d      &   $^2$P$_{3/2}$      & 2.123$-$10  &  2.123$-$10    & 2.71$-$10   \\
  78	& 3s$^2$3p$^4$ ($^1$D)4d      &   $^2$S$_{1/2}$      & 9.430$-$11  &  9.430$-$11    & 1.63$-$10   \\
  79	& 3s$^2$3p$^4$ ($^1$D)4d      &   $^2$G$_{7/2}$      & 3.148$-$10  &  3.148$-$10    &	  $ $	  \\
  80	& 3s$^2$3p$^4$ ($^1$D)4d      &   $^2$G$_{9/2}$      & 3.214$-$10  &  3.214$-$10    &	  $ $	  \\
  81	& 3s$^2$3p$^4$ ($^1$D)4d      &   $^2$P$_{3/2}$      & 1.644$-$10  &  1.644$-$10    & 1.40$-$10   \\
  82	& 3s$^2$3p$^4$ ($^1$D)4d      &   $^2$P$_{1/2}$      & 1.452$-$10  &  1.452$-$10    & 1.12$-$10   \\
  83	& 3s$^2$3p$^4$ ($^1$D)4d      &   $^2$D$_{5/2}$      & 1.731$-$10  &  1.731$-$10    & 2.95$-$10   \\
  84	& 3s$^2$3p$^4$ ($^1$D)4d      &   $^2$D$_{3/2}$      & 1.459$-$10  &  1.459$-$10    & 1.92$-$10   \\
  85	& 3s$^2$3p$^4$ ($^1$D)4d      &   $^2$F$_{5/2}$      & 2.820$-$10  &  2.820$-$10    & 4.94$-$09   \\
  86	& 3s$^2$3p$^4$ ($^1$D)4d      &   $^2$F$_{7/2}$      & 2.929$-$10  &  2.929$-$10    &	  $ $	  \\
  87	& 3s$^2$3p$^4$ ($^1$S)4d      &   $^2$D$_{5/2}$      & 2.648$-$10  &  2.647$-$10    & 1.21$-$09   \\
  88	& 3s$^2$3p$^4$ ($^1$S)4d      &   $^2$D$_{3/2}$      & 2.431$-$10  &  2.430$-$10    & 8.35$-$10   \\
  & & & & & & & & &  \\ \hline				  								   	   
\end{tabular}													   		   
\begin {flushleft}														   
\begin{tabbing}
aaaaaaaaaaaaaaaaaaaaaaaaaaaaaaaaaaaa\= \kill
GRASP3a: Present results from 38  configurations and 3749 levels including only E1 transitions \\
GRASP3b: Present results from 38  configurations and 3749 levels including all E1, E2, M1 and M2 transitions \\
CIV3: Results of Mohan {\em et al} \cite{mm1}  \\ 
\end{tabbing}
\end {flushleft}
\end{document}